\documentclass[11pt]{article}
\pdfoutput=1
\usepackage{jheppub}

\title{Higher Derivative Corrections to Charged Fluids in 2n Dimensions}

\author[a]{Nabamita Banerjee,}
\author[b]{Suvankar Dutta,}
\author[c]{Akash Jain}

\affiliation[a]{Department of Physics, Indian Institute of Science
  Education and Research (IISER), Pune, India}
\affiliation[b]{Department of Physics, Indian Institute of Science
  Education and Research (IISER), Bhopal, India}
\affiliation[c]{Centre for Particle Theory \& Department of
  Mathematical Sciences, Durham University, UK}

\emailAdd{nabamita@iiserpune.ac.in}
\emailAdd{suvankar@iiserb.ac.in}
\emailAdd{akash.jain@durham.ac.uk}

\usepackage{aj-latex/aj-definitions}

\newcommand*{\nfrac}[2]{\genfrac{\langle}{\rangle}{0pt}{}{#1}{#2}}
\usepackage{cancel}
\usepackage{framed}
\definecolor{shadecolor}{gray}{0.9}

\abstract{ We study anomalous charged fluid in $2n$-dimensions ($n\geq 2$) up to sub-leading derivative order. Only the effect of gauge anomaly is important at this order. Using the Euclidean partition function formalism, we find the constraints on different sub-leading order transport
  coefficients appearing in parity-even and odd sectors of the fluid. We introduce a new mechanism to
  count different fluid data at arbitrary derivative order. We show that only the knowledge of 
independent scalar-data is sufficient to find the constraints. In appendix we
  further extend this analysis to obtain fluid data at sub-sub-leading order 
(where both gauge and gravitational anomaly contribute) for
  parity-odd fluid.}

\begin{document}

\maketitle

\section{Introduction and Summary}

In past few years there has been much interest and progress in further
understanding of relativistic, charged, dissipative fluid in presence
of some global anomalies. Presence of quantum anomalies play a crucial
role in transport properties of fluid. The first evidence of
quantum anomaly in fluid transport was holographically observed in \cite{Banerjee:2008th,Erdmenger:2008rm}. The authors found a new parity-odd term (and hence a new transport coefficient) in the charge flavour current. The origin
of this new term can be traced back to gauge Chern-Simons term in the
dual supergravity theory. Soon after these results were published, it
was shown that the new parity-odd term in the charge current is
essential because of the triangle flavour anomalies and the second law of
thermodynamics \cite{Son:2009tf}. In general the second law of thermodynamics
(or equivalently the positivity of divergence of entropy current) imposes
constraints on different transport coefficients. The same
constraint can also be obtained from the equilibrium partition
function of fluid \cite{Banerjee:2012iz,Jensen:2012jh}. Equilibrium partition function provides an
alternate and a microscopically more transparent way to derive the
constraints on these transport coefficients. A generalization of this approach
for charged $U(1)$ anomalous fluid in arbitrary even dimensions up to leading order has been considered in \cite{Banerjee:2012cr}.

In \cite{Bhattacharyya:2013ida} Bhattacharyya $et. al.$ studied parity odd 
transport for a four dimensional non-conformal charged fluid at second
order in derivative expansion. In four spacetime dimensions the effect of anomaly
appears at one derivative order and the parity-odd transport
coefficients at this order are determined in terms of anomaly
coefficient. In this paper the authors studied the transport
properties at second order and found that out of 27
transport coefficients 7 are fixed in terms of anomaly and lower order
transport coefficients. The goal of our current paper is to generalize
this work to arbitrary even dimensions. In $2n$ spacetime dimensions
the leading effect of anomaly appears at $(n-1)$ derivative
order. Hence the subleading corrections appear at $n$th derivative
order. The aim of this paper is to study the constraints on transport
coefficients appearing at subleading order. We
innovate a systematic mechanism to compute different fluid data at
arbitrary derivative order (parity odd or even). We list all possible scalars,
vectors and tensors at any arbitrary derivative order in this
paper. It seems to be rather difficult to find the independent sets.
However, we argue that it is possible to get the correct constraint
relations between transport coefficients even without knowing the
independent sets of fluid data.

Our analysis is not valid in two spacetime dimensions. In two
dimensions the parity odd terms appear at zero derivative order
itself, and hence parity-odd and parity even sectors are not
independent at any arbitrary order. Independence of these two sectors
is important in our computation.

In the parity-even sector, the leading correction appears at first
order in derivative expansion, $e.g.$ shear viscosity and bulk
viscosity terms in energy momentum tensor etc. In this paper we have
extended our calculation to include the sub-leading order correction
($i.e.$ second order corrections) to parity-even sector in
constitutive relations in arbitrary even dimensions in presence of
$U(1)$ gauge anomaly. This completes the description of fluid dynamics
up to sub-leading order in derivative expansion (both in parity-odd
and even sectors) in arbitrary even dimensions with $abelian \ gauge \
anomaly$.

The organization of our paper is as following. In \cref{sec:scheme}
we explain our notation and perturbation scheme which we
use in this paper. In \cref{Sec:euclidean} we construct the
partition function for both gauge invariant and non-invariant sectors
and compute the constitutive relations from the partition
function. We also describe the construction of the anomalous entropy current.
\Cref{sec:counting} is the most important section of
this paper. Here we first describe how to construct fluid data at arbitrary 
derivative order. Next, we list all the leading and sub-leading order
scalars, vectors and tensors which may appear in constitutive
relations up to sub-leading order in derivative expansion both in
parity-even and odd sectors. Although, we have not been able to find
the \emph{`independent'} parity-odd vectors and tensors at sub-leading order,
this does not inhibit us from finding the constraints on the transport
coefficients. We elaborate this issue in
\cref{sec:basisdata}. Finally, in \cref{sec:fluidrelations} we
list the constraint on the transport coefficients up to sub-leading
order. In appendices we explain the Kaluza-Klien
decomposition (\cref{apn:KK}) and sub-sub-leading order counting
(\cref{apn:subsubleading}).

\section{Scheme and the Perturbative Expansion} \label{sec:scheme}

We consider a $2n$-dimensional spacetime manifold $\cM_{(2n)}$ with
metric $\df s^2 = G_{\mu\nu}\df x^\mu \df x^\nu$ and gauge field 1-form $\cA = \cA_\mu \df x^\mu$. We want to study fluid
dynamics in this background. A fluid is a statistical system in local
thermodynamic equilibrium, which is generally characterized in terms
of (covariant) \emph{energy-momentum tensor} $\bar\cT^{\mu\nu}$,
(covariant) \emph{charge current} $\bar\cJ^{\mu}$ and their
constitutive equations
\bea{\label{E:covariantCons}
	\hat\N_\mu\bar\cT^{\mu\nu} 
	&= \cF^{\nu\r} \bar\cJ_\r + \underaccent{\sim}{\fT}^{\nu}, \qquad
	\hat\N_\mu \bar \cJ^\mu = \undertilde{\fJ}.
}

$\cF = \df \cA$ is field strength for $\cA$. Here we have introduced a $U(1)$ anomaly $\undertilde{\fJ}$ and a gravitational anomaly $\underaccent{\sim}{\fT}^{\nu}$. The form of these anomalies is well known in literature \cite{Jensen:2013kka}. Most of our work here will be concentrated on fluid upto subleading derivative order, where only $U(1)$ anomalies contribute: 
\bee{\label{E:anomalouscurrent}
	\undertilde\fJ = (n+1)  C^{(2n)} \star \cF^{\wedge n} = (n+1)  C^{(2n)} \frac{1}{2^n}\epsilon^{\mu_1\nu_1 \cdots \mu_n\nu_n}
  \cF_{\mu_1\nu_1}\ldots \cF_{\mu_n\nu_n}.
}
$\underaccent{\sim}{\fT}^{\nu}$ only starts getting values at subsubleading derivative order. Let us explain our notation here. 
\begin{itemize}
\item All the fluid quantities (like currents, transport coefficients,
  independent terms etc.) appearing in parity-odd sector, are denoted
  by `tilde' (e.g. $\tilde A$). On the other hand we use no
  special notation for parity-even sector (e.g. $A$). Wherever
  applicable, $\bar A = A + \tilde A$ denotes the total quantity
  (parity-odd and parity-even).
\item $\hat\N$ and $\N$ denote the covariant derivative and on $\cM_{(2n)}$ and the equilibrium manifold $\cM_{(d-1)}$ respectively. We use $\wedge$ and $\star$ as wedge product and Hodge Dual on all manifolds, as no confusion is possible.
\end{itemize}

Due to dissipative nature of fluid, it is not possible to write an
exact generating functional $W$ (or action) for fluids from which
one can derive the energy-momentum tensor $\bar\cT^{\mu\nu}$ and
charge current $\bar{\cJ^{\mu}}$. Therefore we write their most
generic forms, allowed by symmetries, in terms of fundamental fluid
variables and their derivatives in a particular thermodynamic
ensemble. In our analysis we consider the fluid variables to be
\emph{temperature} $\vq$, \emph{chemical potential}\footnote{Actually
  $\nu = \mu/\vq$, where $\mu$ is the chemical potential.} $\nu$ and
\emph{fluid four-velocity} $u^\mu$ with $u^\mu u_\mu = -1$.

We prefer to work in \emph{Landau Frame}, where 
all the dissipation terms are transverse to the direction of the fluid
flow. Hence, we can decompose $\bar\cT^{\mu\nu}$ and $\bar{\cJ^{\mu}}$
as
 \bee{\label{E:TJdefi} \bar\cT^{\mu\nu} = E(\vq,\nu) u^\mu u^\nu +
  \bar\Pi^{\mu\nu}, \quad \bar\cJ^{\mu} = Q(\vq,\nu) u^\mu +
  \bar\U^{\mu}, } 
where $\bar\Pi^{\mu\nu}$ and $\bar\U^\mu$ are the most generic
symmetric tensor and vector made out of fluid variables. In the Landau
frame
 \bee{ \label{E:LGC} u_\mu \bar\Pi^{\mu\nu} = 0, \qquad
  u_\mu \bar\U^{\mu} = 0.  }
The easiest way to implement this is to project all vectors or tensors
appearing in $\bar\U^{\mu}$ or $\bar\Pi^{\mu\nu}$, transverse to
$u^\mu$ using the projection operator 
$$P^{\mu\nu} = G^{\mu\nu} +
u^\mu u^\nu.$$ 

Since fluid is a low energy fluctuation about the local thermodynamic
equilibrium, $\bar\Pi^{\mu\nu}$ and $\bar\U^\mu$ can be expanded in
derivatives of fundamental fluid variables ($\vq, \nu, u^\mu$):
\bee{\label{E:FirstDerExp} \bar\Pi^{\mu\nu} = \bar\Pi^{\mu\nu}_{(0)} +
  \bar\Pi^{\mu\nu}_{(1)} + \bar\Pi^{\mu\nu}_{(2)} \ldots, \qquad
  \bar\U^{\mu} = \bar\U^{\mu}_{(0)} + \bar\U^{\mu}_{(1)} +
  \bar\U^{\mu}_{(2)} \ldots, } 
where $\bar\Pi^{\mu\nu}_{(N)}$ and
$\bar\U^{\mu}_{(N)}$ involves $N$ number of derivatives on fluid
variables.  The terms on RHS can have the most generic form as,
\bea{\label{E:ConstitutiveRelations} \bar\Pi^{\mu\nu}_{(N)} &=
  \sum_t\t_{(N)t} (\vq,\nu)\mathbf T_{(N)t}^{\mu\nu} +
  P^{\mu\nu}\sum_t\s_{(N)t}(\vq,\nu) \mathbf S_{(N)t}, \nn\\ 
  \bar\U^\mu_{(N)} &= \frac{\vq_o}{\e+
    P_o}\sum_t\nu_{(N)t}(\vq,\nu)\mathbf V_{(N)t}^{\mu}, }
where $\mathbf S_{(N)t}$, $\mathbf V^{\mu}_{(N)t}$ and $\mathbf
T^{\mu\nu}_{(N)t}$ are a collection of all possible gauge invariant
scalars, vectors and symmetric traceless tensors (collectively known
as \emph{data}) respectively, made out of fluid variables and source
fields at $N$ derivative order. $\sum_t$ corresponds to sum over
independent terms at any particular derivative order. The data which
is required for our computation has been enlisted in
\cref{sec:counting}.

In \cref{E:ConstitutiveRelations}, the expression for
$\bar\Pi^{\mu\nu}_{(N)}$ and $\bar\U^\mu_{(N)} $ are fixed up to some
undetermined coefficients appearing at each derivative order. Therefore,
a fluid is characterized by an infinite set of such unknown functions
($\t_{(N)t},\s_{(N)t},\nu_{(N)t}$), known as \emph{transport
  coefficients}. Fluid up to a particular derivative order is
characterized by a finite number of such transport coefficients. In
general, these transport coefficients are not all independent. The
second law of thermodynamics (or equivalently, positivity of local
entropy current) imposes restrictions on different transport
coefficients\footnote{Similar restrictions are also applicable to non-relativistic fluids and has recently been addressed
for charged non-relativistic fluids in \cite{Banerjee:2014mka}.} \cite{LandauBook}.  
Such relations among various transport
coefficients are known as \emph{constraints}.

\cite{Banerjee:2012iz} uses a different mechanism to find `some' of
these constraints.
The idea is to write an equilibrium partition function for the fluid
and derive the energy-momentum tensor and charge current from that
partition function. Because of dissipation it is not possible to write
a generating functional ($W$) for the fluid. However, one can still
write a generating functional in equilibrium configuration, which we
denote by $W^{eqb}$. Using $W^{eqb}$ one can find all the
constraint relations involving transport coefficients which comes with
data that survives at equilibrium.

More precisely, if the theory has a timelike Killing vector $\o^\mu$,
we can write an Euclidean generating functional using the background
fields and Killing equation on the decomposed manifold
$S^1\times\cM_{(d-1)}$. Here $S^1$ is the euclidean time circle along
$\o^\mu$ with time period $\tilde\b$, and $\cM_{(d-1)}$ is the spacetime
transverse to $\o^\mu$. \cite{Banerjee:2012iz} has conveniently chosen
$\o^\mu = \dow_0$. Therefore, one can decompose the background in Kaluza-Klein form,
\ben
	\df s^2 
	&=& G_{\mu\nu}\df x^\mu \df x^\nu
	= -\E{2\s} \lb \df t + a_i \df x^i \rb^2
	+ g_{ij}\df x^i \df x^j, \nn \\
{\cal A} &=& A (dt +a_idx^i) + A_i dx^i. \label{E:background}
\een

For more details please refer \cref{apn:KK}. Using this choice along with the Landau Gauge
conditions and velocity normalization, the most-generic
energy-momentum tensor and charge current in \cref{E:TJdefi} on $\cM_{(d)}$
can be decomposed into scalars, vectors and tensors on
$S^1\times\cM_{(d-1)}$:
\bea{\label{E:CurrentsGeneric}
  \bar T^{ij}	&= E (\vq,\nu) v^i v^j + \bar\pi^{ij}, \nn\\
  \bar T^{i} 	&= - \E{\s} \lb E(\vq,\nu) v^i \sqrt{1+ v_i v^i} +
  \frac{v_j \bar\pi^{ij}}{\sqrt{1+ v_i v^i}} \rb, \nn\\ 
  \bar T 		&= \E{2\s} \lb E(\vq,\nu) (1+ v_i v^i) +
  \frac{v_i v_j \bar\pi^{ij}}{(1+ v_i v^i)} \rb, \nn \\ 
  \bar J^{i} 	&= Q(\vq,\nu) v^i + \bar\vs^{i},\nn\\
  \bar J &= - \E{\s} \lb Q(\vq,\nu) \sqrt{1+ v_i v^i} + \frac{v_i
    \bar\vs^{i}}{\sqrt{1+ v_i v^i}} \rb, }
where
\bee{\nn
   \bar T = \bar\cT_{00},\ \bar T^i =
  \bar\cT^i_{\ 0}, \ \bar T^{ij} = \bar\cT^{ij}; \quad
  \bar J = \bar\cJ_0, \ \bar J^i = \bar\cJ^i, }
and 
\bee{\nn \bar\pi =\bar\Pi_{00} ,  \ \bar\pi^i =
  \bar\Pi^i_{\ 0}, \ \bar\pi^{ij} = \bar\Pi^{ij}, \quad
  \bar\vs = \bar\U_0, \ \bar\vs^i = \bar\U^i, \quad v = u_0, v^i = u^i.}

Indices on $\cM_{(d-1)}$ are raised and lowered using
$g^{ij}$. Details of Kaluza-Klein decomposition of fluid variables and
background fields have been given in \cref{apn:KK}. 

Since the fluid we are considering is in local thermodynamic
equilibrium, we can write the fluid variables as a spatial derivative
expansion about their equilibrium values
\bea{\label{E:fluidvarExp}
  \vq &= \vq_o + \bar\D^{(1)}\vq + \bar\D^{(2)}\vq + \ldots \nn \\
  \nu &= \nu_o + \bar\D^{(1)}\nu + \bar\D^{(2)}\nu + \ldots \nn \\ 
  v^i &= v^i_o + \bar\D^{(1)}v^i + \bar\D^{(2)}v^i + \ldots .
 }
 The terms with subscript `$o$' are the equilibrium values, while
 $\bar\D^{(N)}$ designates the $N^{th}$ derivative
 corrections\footnote{In this paper, $\tilde\D^{(n)} A$ denotes parity-odd
   $n^{th}$ derivative corrections to a fluid quantity $A$, while $\D^{(n)} A$
   represents the parity-even $n^{th}$ derivative corrections. Entire
   derivative correction is denoted by $\bar\D^{(n)} A = \D^{(n)} A + \tilde\D^{(n)}
   A$. 
   }. The zeroth component of fluid velocity $u_0=v$ also gets
 derivative corrections which are determined by the corrections to
 $v^i$ using the four-velocity normalization.
 Similarly all the transport coefficients can also be expanded using
 the Taylor Series expansion
 \bee{ \a(\vq,\nu) = \a_o(\vq_o,\nu_o) +
   \bar\D^{(1)} \a+\bar\D^{(2)} \a+\ldots.  }

 Therefore the energy-momentum tensor and charge current receive two
 fold derivative corrections. First of all we write these expressions
 as a derivative expansion in terms of fluid variables in
 \cref{E:FirstDerExp}. Secondly, each term in that expansion can be
 further expanded around the equilibrium values of fluid variables
 according to \cref{E:fluidvarExp}. Thus we finally get
\bea{
   \bar\pi^{ij} &= \lB \bar \pi^{ij}_{o(0)} \rB + \lB \bar\D^{(1)} \bar
   \pi^{ij}_{(0)} + \bar\pi^{ij}_{o(1)} \rB + \lB \bar\D^{(2)} \bar
   \pi^{ij}_{(0)} + \bar\D^{(1)} \bar \pi^{ij}_{(1)} +
   \bar\pi^{ij}_{o(2)}
   \rB\ldots, \nn\\
	\bar \vs^{i} &= 
	\lB
		\bar \vs^{i}_{o(0)}
	\rB 
	+ \lB
		\bar\D^{(1)}\bar \vs^{i}_{(0)}
		+ \bar \vs^{i}_{o(1)}
	\rB
	+ \lB
		\bar\D^{(2)}\bar \vs^{i}_{(0)}
		+ \bar\D^{(1)}\bar\vs^{i}_{(1)}
		+ \bar\vs^{i}_{o(2)}
	\rB\ldots.
}
Expansion of time components can be determined from these using Landau
gauge condition \cref{E:LGC}. 

We choose the equilibrium convention for $\vq$ and $\nu$ by
identifying their equilibrium values to be the red-shifted temperature
and Wilson loop
in the lower dimensional theory
 \bee{ \frac{1}{\vq_o} = \b_o =
  \tilde\b\sqrt{-G_{00}} = \tilde\b\E{\s}, \qquad \nu_o = \tilde\b
  \cA_0.  }

In the next section we construct the equilibrium partition function and
obtain energy-momentum tensor and charge current in terms of background
data following \cite{Banerjee:2012iz}. After that, we compare these
stress tensor and current with the fluid stress tensor and current
order by order in derivative expansion to find the constraints among
the transport coefficients at any particular derivative order. A
typical constraint will connect transport coefficients at equilibrium
$\lbr\a_o(\vq_o,\nu_o)\rbr$ and their derivatives with respect to
$\vq_o$ and $\nu_o$ (up to a particular derivative order)
 \bee{
  \fC\Big( \lbr\a_o(\vq_o,\nu_o)\rbr, \lbr\dow\a_o(\vq_o,\nu_o)\rbr
  \Big) = 0.  }
We can extrapolate this constraint to non-equilibrium configurations:
\bee{
	\fC\Big( \lbr\a(\vq,\nu)\rbr, \lbr\dow\a(\vq,\nu)\rbr \Big) = 0,
}
while doing this, we are making an error of at least one derivative
order higher, which will be compensated at next derivative order
computation. This is how we find the generic constraints among fluid
transport coefficients.  Please note that while the equality
constraints determined by this procedure are generic, the inequality
constraints are not determined by this method.

\section{Equilibrium Partition Function} \label{Sec:euclidean}

The equilibrium partition function\footnote{The partition function may be thought of as the
  Euclidean action for the fluid living on the background with
  coordinate time $t$ compactified on a circle of length $\tilde\beta$} $W^{eqb}$ of the theory can generally be disintegrated into two parts:
\begin{equation}
 W^{eqb}= W^{eqb}_{(C)}+W^{eqb}_{(A)}.
\end{equation}

The first one is the `conserved' partition function which is gauge and diffeomorphism invariant, and generates conserved part of currents denoted by $\bar\cT^{\mu\nu}_{(C)}, \bar\cJ^\mu_{(C)}$. The other piece is not gauge-invariant and is referred to be `anomalous' partition function. It generates anomalous piece of `consistent currents' which will not be gauge-invariant in general. By defining a consistent subtraction scheme (Bardeen-Zumino currents), we can make these anomalous currents gauge invariant (see \cite{Jensen:2012kj} for details) which we denote by $\bar\cT^{\mu\nu}_{(A)}, \bar\cJ^\mu_{(A)}$. Their value at equilibrium is fixed by anomaly, and upto subleading order is given by:
\bea{
	\bar\cT^{\mu\nu}_{(A)} &= - 2 C^{(2n)} \sum_{m=1}^{n} \ ^{n+1}\bC_{m+1} \vq_o^2 \nu_o^{m+1} \star\lb u_o \wedge \cX_{o1}^{\wedge (m-1)} \wedge \cX_{o2}^{\wedge(n-m)} \rb^{(\mu} u_o^{\nu)} \nn \\
	&\equiv - 2 C^{(2n)} \sum_{m=1}^{n} \ ^{n+1}\bC_{m+1} \vq_o^2 \nu_o^{m+1} l_{o(m)}^{(\mu} u_o^{\nu)}, \label{E:Tmunu_A} \\
	\bar\cJ^\mu_{(A)} &= - C^{(2n)} \sum_{m=1}^{n} (n+1) \ ^{n}\bC_{m} \vq_o \nu_o^m \star\lb u_o \wedge \cX_{o1}^{\wedge(m-1)} \wedge \cX_{o2}^{\wedge (n-m)} \rb^\mu \nn \\
	&\equiv - C^{(2n)} \sum_{m=1}^{n} (n+1) \ ^{n}\bC_{m} \vq_o \nu_o^m l^\mu_{o(m)}. \label{E:Jmu_A}
}

Here $\lbr \cX_1, \cX_2\rbr$ are $\lbr - \vq \df u, \df \cA + \vq\nu \df u \rbr$ projected transverse to $u^\mu$, and their equilibrium values upon KK reduction reduce to: $\lbr f_1, f_2\rbr = \lbr \tilde\vq \df a, \df A \rbr$. Hence we have\footnote{To get these and some further results we have to used the ideal order results $v_o = -\E{\s}, v^i_o = 0$, which we will derive in \cref{sec:fluidrelations}. We use it here to simplify the notation.}:
\bea{
	\bar\cT^{i}_{(A)} &= C^{(2n)} \sum_{m=1}^{n} \ ^{n+1}\bC_{m+1} \E{\s} \vq_o^2 \nu_o^{m+1} l_{o(m)}^{i}, \qquad \bar\cT^{ij}_{(A)} = \bar\cT_{(A)} = 0, \label{E:T_A}  \\
	\bar J^i_{(A)} &= - C^{(2n)} \sum_{m=1}^{n} (n+1) \ ^{n}\bC_{m} \vq_o \nu_o^m l^i_{o(m)}, \qquad \bar J_{(A)} = 0. \label{E:J_A}
}

Let us now concentrate on $W^{eqb}_{(C)}$. It's variation on background (\ref{E:background}) will determine the conserved currents:
\be \d  W^{eqb}_{(C)} = \int d^{2n} x \ \sqrt{G}\left[ -\half
  \bar\cT^{\mu\nu}_{(C)} \delta G_{\mu\nu} + \bar\cJ^{\mu}_{(C)} \delta
  {\cA_{\mu}}\right]. \ee
And hence,
\bee{\label{E:consistentDefn}
	\bar\cT^{\mu\nu}_{(C)} = 2\frac{\d W^{eqb}_{(C)}}{\d G_{\mu\nu}}, \qquad
	\bar\cJ^{\mu}_{(C)} = \frac{\d W^{eqb}_{(C)}}{\d \cA_{\mu}}.
}

Kaluza-Klein decomposition of \cref{E:consistentDefn} gives,
 \bee{ \bar T^{ij}_{(C)} =
  2\vq_o \frac{ \d W^{eqb}_{(C)}}{\d g_{ij}}, \qquad
	\bar T^{i}_{(C)}
	+ \E{\s}\vq_o\nu_o \bar\cJ^i_{(C)}
	= 
	\vq_o \frac{ \d W^{eqb}_{(C)}}{\d a_i}, \qquad
	\bar T_{(C)}
	= 
	\E{2\s}\vq_o^2 \frac{ \d W^{eqb}_{(C)}}{\d \vq_o}, \nn
}
\bee{\label{E:identifications}
	\bar J^i_{(C)} = \vq_o \frac{ \d W^{eqb}_{(C)}}{\d A_i}, \qquad
	\bar J_{(C)}
	= 
	- \E{\s} \frac{\d W^{eqb}_{(C)}}{\d \nu_o}.
}
Here we have switched the basis to $\vq_o = \E{-\s}/\tilde\b$ and
$\nu_o = \tilde\b A$ for later convenience. $a^i$ is the Kaluza-Klein
gauge field. Note that while $W^{eqb}_{(C)}$ is gauge invariant, its integrand does not need to be. We can include a typical Chern-Simons term to it, which is defined such that its integral is gauge invariant\footnote{We have left the terms in $I^{2n-1}$ which can be related to others upto a total derivative.}:
\bee{\label{E:lowerCS}
	\int_{\cM_{(2n-1)}} I^{2n-1}
	=
	- \int d^{2n-1}x \sqrt{g} \lbr \sum_{m=1}^{n}
   \ ^{n}\bC_{m-1} C_{m-1} A_i l^i_{o(m)} + \tilde\vq C_{n}
   a_{i}l^i_{o(n)} \rbr.
}
Here $C_m$'s are constants. This is indeed a valid Chern-Simons form as at equilibrium $l^i_{o(m)}$ is just made of Chern classes of $f_1$ and $f_2$:
\bee{
	l^i_{o(m)} = \star\lb f_1^{\wedge(m-1)} \wedge f_2^{\wedge (n-m)} \rb^i.
}

For the gauge-invariant integrand, we assume that curvature
scales of the background $\cM_{(d-1)}$ is much much larger than the
mean free path of the fluid, therefore the whole manifold can be
thought of as union of various flat patches. The system can be thought
of in thermal equilibrium in each local patch. On each patch we can
define the euclidean partition function locally, hence giving us
\bee{
	W^{eqb}_{(C)} = \int d^{2n-1}x \sqrt{g} \ \b(\vec x) P(\vec x) + \int d^{2n-1}x \sqrt{g} \star I^{2n-1},
}
where $P(\vec x)$ is local thermodynamic \emph{pressure} and $\b(\vec
x)$ is local thermodynamic temperature. Given pressure, we can use the
thermodynamic relations in local patch
 \bee{ \df P = \frac{\e +
    P}{\vq} \df \vq + \vq q\df \nu, \quad \e+P= \vq s + \nu q, }
to define \emph{energy density} $\e$, \emph{entropy density} $s$ and \emph{charge density} $q$ of
the fluid. All are functions of $\vq$ and $\nu$. We can expand
$W^{eqb}_{(C)}$ around its equilibrium value as
 \bee{ W^{eqb}_{(C)} =
  \int d^{2n-1}x  \sqrt{g} \ \b_o P_o + \bar\D W^{eqb}_{(C)}.  }
 Derivative correction to the ideal fluid partition function is
 denoted by $\bar\D W^{eqb}_{(C)}$, which will contain all the possible gauge invariant scalars
 made out of background metric and gauge
 field components at a particular derivative order. We have computed these scalars (till the derivative
 level of our interest) in \cref{sec:counting}.
 
Collating together the conserved currents in \cref{E:identifications} and the anomalous pieces in \cref{E:J_A,E:T_A}, and varying the Chern-Simons terms in $W_{(C)}^{eqb}$ i.e. \cref{E:lowerCS}, we can finally write:
\bee{
	\bar T^{ij}
	= 
	2\vq_o \frac{\d W_{(C)}^{eqb}}{\d g_{ij}}, \nn
}
\bee{
	\bar T^{i}
	+ \E{\s}\vq_o\nu_o \bar J^i
	= 
	\vq_o \frac{ \d W_{(C)}^{eqb}}{\d a_i}
	- \vq_o\sum_{m=1}^{n} \ ^{n-1}\mathbf C_{m-1} \lbr 
		\frac{n(n+1)}{(m+1)} C^{(2n)} \E{\s}\vq_o\nu_o \nu_o^{m} l^i_{o(m)}
		+ n C_{m} \tilde\vq l^i_{o(m)}
	\rbr, \nn
}
\bee{
	\bar T
	= 
	\E{2\s}\vq_o^2 \frac{ \d W_{(C)}^{eqb}}{\d \vq_o}, \nn
}
\bee{
	\bar J^i = \vq_o \frac{ \d W_{(C)}^{eqb}}{\d A_i} 
	- \vq_o\sum_{m=1}^{n} \ ^{n-1}\mathbf C_{m-1} \lbr 
		\frac{n(n+1)}{m} C^{(2n)} \nu_o^{m} l^i_{o(m)}
		+ n C_{m-1} l^i_{o(m)}
	\rbr, \nn
}
\bee{ \label{E:parti_perturb}
	\bar J
	= 
	- \E{\s} \frac{\d W_{(C)}^{eqb}}{\d \nu_o}.
}

Comparing these to the most generic fluid expressions in \cref{E:CurrentsGeneric} we can compute the
constraints. Thus, we see that it is only the gauge invariant
part $W_{(C)}^{eqb}$ of the partition function that we need to
evaluate at any desired order.

\subsection{Anomalous Entropy Current}

In last section we reviewed a procedure to get equality type constraints among fluid transport coefficients. It is generally 
known that these very constraints can also be get by demanding existence of an entropy current whose divergence is positive 
semi-definite. The most generic Entropy Current can be written as:
\bee{
	\bar \cJ^\mu_S = \bar \cJ^\mu_{S(C)} + \bar \cJ^\mu_{S(A)},
}
where $\bar \cJ^\mu_{S(A)}$ is the part which captures the explicit dependence on anomaly coefficients. However, 
the other piece $\bar \cJ^\mu_{S(C)}$ can get implicit dependence on the anomaly coefficients through the fluid equations
of motion. We need to demand this current to be positive 
semi-definite,
\bee{
	\hat\N_\mu\bar \cJ^\mu_S = \hat\N_\mu\bar \cJ^\mu_{S(C)} + \hat\N_\mu\bar \cJ^\mu_{S(A)} \geq 0,
}
whenever EOM are satisfied. For equilibrium fluid configuration, both the pieces can be demanded to be positive semi-definite
 separately. Such decoupling is not always possible, as the fluid equations of motion depend on anomaly coefficients, which can induce some implicit anomaly dependence in $\bar \cJ^\mu_{S(C)}$. 
However, for equilibrium fluid
configurations, the equations of motion are trivially satisfied and thus entire information of anomaly can be incorporated in 
$\bar \cJ^\mu_{S(A)}$.
 Hence, if any part of $\hat\N_\mu\bar \cJ^\mu_{S(C)}$ couple to $\hat\N_\mu\bar \cJ^\mu_{S(A)}$, 
the respective transport coefficients will be determined in terms of anomaly coefficients, and hence will 
be present in $\bar\cJ_{S(A)}^\mu$ at the first place. Therefore all the information about constraints
 among fluid transport coefficients is encoded in the existence of $\cJ^\mu_{S(C)}$. In 
\cite{Bhattacharyya:2013lha,Bhattacharyya:2014bha} the author 
 gives an explicit construction of entropy current from Eqb. Partition Function.

Now concentrating on the second term: at equilibrium, $\hat\N_\mu\bar \cJ^\mu_{S(A)} \geq 0$, since it does not have any independent 
coefficients, just constants, one cannot apply any constraints for it to be satisfied. Therefore $\cJ^\mu_{S(A)}$ must 
be exact. But any current is always ambiguous upto some exact terms, and hence we can choose $\cJ^\mu_{S(A)} = 0$ equally well. 
We can hence write in a generic hydrodynamic frame\footnote{We have used the thermodynamic functions $\e,q,s$ here, which will
be explicitly proved in \cref{sec:fluidrelations}.}:
\bea{
	\bar\cT^{\mu\nu}_{(A)} &
	= 2 \e u^{(\mu}_{(A)} u^{\nu)}_{(C)} + 2 \bar q^{(\mu}_{(A)} u^{\nu)}_{(C)} + 
2 \bar q^{(\mu}_{(C)} u^{\nu)}_{(A)} + \tilde\Pi^{\mu\nu}_{(A)}, \\
	\bar\cJ^{\mu}_{(A)} &= q u^{\mu}_{(A)} + \bar\U^\mu_{(A)}, \\
	0 &= s u^{\mu}_{(A)} + \bar\U^\mu_{S(A)}.
}

Note that in the expression for $\bar\cT^{\mu\nu}_{(A)}$ we have used the fact that anomalies are 
parity-odd. Now depending on the choice of hydrodynamic frame, these conditions can be used to determine 
anomalous dissipative parts of the various currents. For example, if we define $u^\mu$ such that it does not 
contribute to anomaly, i.e. $u^\mu_{(A)} = 0$, we will get:
\bee{
	\tilde\U^{\mu}_{(A)} = \bar\cJ^\mu_{(A)}, \qquad
	q^\mu_{(A)} = - \bar\cT^{\mu\nu}_{(A)} u_{\nu(C)}, \qquad
	\tilde\Pi^{\mu\nu}_{(A)} = 2\bar\cT^{(\mu\a}_{(A)} \lb \d^{\nu)}_{\ \a} + u^{\nu)}_{(C)}u_{\a(C)} \rb, \qquad
	\tilde\U^{\mu}_{S(A)} = 0.
}
This is the neatest frame for anomalies. Similar results for $U(1)$ anomaly were derived in \cite{Loganayagam:2011mu}, 
however these expressions are also applicable to gravitational anomalies\footnote{restricted to equilibrium configurations}. 
Here we present explicit expressions for the anomalous parts of currents, in presence of both $U(1)$ 
and gravitational anomaly. Following the generic expressions given in \cite{Jensen:2013kka}, these can be computed 
directly from the anomaly
polynomial. The anomaly polynomial in $2n$ dimensions up to $(n+1)$ derivative order is given as \cite{Loganayagam:2012pz},
\begin{equation}
 {\cal{P}}= C^{(2n)} \cF^{\wedge (n+1)} + c_m \cF^{\wedge (n-1)} \wedge \mathrm{Tr}[\fR\wedge \fR], 
\end{equation}
where, $C^{(2n)}$ is gauge anomaly coefficient which we have already introduced in the last section
 and $c_m$ is gravitational anomaly coefficient. The two form $\fR$ is defined in terms of the Riemann tensor as,
\begin{equation}
 \fR^\a_{\ \b} = \cR^{\a}_{\ \b\g\d} \df x^\g\wedge \df x^\d.
\end{equation}
Taking appropriate derivative of the above, one can find explicit expressions for anomalous parts of the currents. The leading 
part of the currents proportional to the gauge anomaly coefficient $C^{(2n)}$ have already been given in \cref{E:Tmunu_A,E:Jmu_A}.
Here we present the subleading order contributions to currents coming due to the gravitational anomaly,
\bem{
	\bar\cJ^\mu_{(A)} 
	= 
	c_m(n-1) \lB
		\star\lb u_o \wedge \cF^{\wedge(n-2)} \wedge \L_{\a\b} \lb \L^{\a\b} U - 2\fR^{\a\b} \rb \rb^\mu \dbrk
		+ \sum_{m=1}^{n-2} \ ^{n-2}\bC_m (\vq_o\nu_o)^m 
		\star\lb
			u_o \wedge U^{\wedge(m-1)} \wedge \cF^{\wedge(n-2-m)} \wedge \lb\fR^{\a\b} - 
\L^{\a\b} U \rb \wedge \lb\fR_{\a\b} - \L_{\a\b} U \rb
		\rb^\mu
	\rB,
}
where,
\bee{
	\L_{\mu\nu}
	=
	\half \lb
		U_{\mu\nu} 
		- 4\frac{1}{\vq_o}u_{o[\nu}P_{\mu]\a}\hat\N^{\a}\vq_o
	\rb, \quad 
	U_{\mu\nu}= 2 P_{[\mu\a}P_{\nu]\b} \hat \nabla^{\a}u^{\b}.
}

The heat current has the form,
\bem{
	\fq^{\mu}_{(A)} 
		= 
	- c_m \frac{1}{\vq_o}  \lB
		\star\lb u_o \wedge\cF^{\wedge(n-1)}\rb^{\mu} \L^{\a\b}\L_{\a\b} \dbrk
		+ \sum_{m=2}^{n-1} \ ^{n-1}\bC_{m} (m-1) (\vq_o\nu_o)^m 
		\star\lb
			u_o \wedge U^{\wedge(m-2)} \wedge \cF^{\wedge(n-1-m)} \wedge \lb\fR^{\a\b} - \L^{\a\b} U \rb \wedge \lb\fR_{\a\b} - \L_{\a\b} U \rb
		\rb^{\mu} \dbrk
		- 2 \sum_{m=1}^{n-1} \ ^{n-1}\bC_{m} (\vq_o\nu_o)^m 
		\star\lb
			u_o\wedge U^{\wedge(m-1)} \wedge \cF^{\wedge(n-1-m)} \wedge \L^{\a\b} \lb\fR_{\a\b} - \L_{\a\b} U \rb
		\rb^{\mu}
	\rB .
}
Finally, the stress tensor looks like,
\bem{
	\bar\cT^{\mu\nu}_{(A)}
	=
	4c_m\hat\N_\r\lB
		\sum_{m=1}^{n-1} \ ^{n-1}\bC_m (\vq_o\nu_o)^m 
		\star\lb
			u_o \wedge U^{\wedge(m-1)}\wedge \cF^{\wedge(n-m-1)} \wedge \lb \fR^{\r(\nu} - \L^{\r(\nu} U \rb
		\rb^{\mu)} \dbrk
		- \star\lb u_o \wedge \cF^{\wedge(n-1)} \L^{\r(\nu}\rb^{\mu)}
	\rB
	- 2 \vq_o u^{(\mu} \fq^{\nu)}_{(A)}.
}

 
 Instead if we are working in Landau Frame, 
where $\bar q^{\mu}_{(A)} = \bar q^{\mu}_{(C)} = 0$, we will get condition:
\bee{
	- \bar\cT^{\mu\nu}_{(A)} u_{\nu(C)} 
	= \lb \e G^{\mu\nu} + \bar\cT^{\mu\nu}_{(C)} \rb u_{\nu(A)}
	= \lb \e G^{\mu\nu} + \bar\Pi^{\mu\nu}_{(C)} \rb u_{\nu(A)}.
}

We need to invert $\lb \e G^{\mu\nu} + \bar\Pi^{\mu\nu}_{(C)} \rb$, which can be done perturbatively in derivatives. 
To leading order:
\bee{
	u^\mu_{(A)} = -\frac{1}{\e+P} \bar\cT^{\mu\nu}_{(A)} u_{\nu(C)} + \ldots,
}
and hence 
\bee{
	\tilde\U^{\mu}_{(A)} = \bar\cJ^\mu_{(A)} + \frac{q}{\e+P} \bar\cT^{\mu\nu}_{(A)} u_{\nu(C)} + \ldots, \qquad
	\tilde\Pi^{\mu\nu}_{(A)} = 2\bar\cT^{(\mu\a}_{(A)} \lb \d^{\nu)}_{\ \a} + \frac{\e}{\e+P} u^{\nu)}_{(C)}u_{\a(C)} \rb + \ldots,
}
\bee{
	\tilde\U^{\mu}_{S(A)} 
	= \frac{s}{\e+P} \bar\cT^{\mu\nu}_{(A)} u_{\nu(C)} + \ldots
	= \frac{1}{\vq} \bar\cT^{\mu\nu}_{(A)} u_{\nu(C)} + \nu\bar\cJ^\mu_{(A)}
	- \nu \tilde\U^{\mu}_{(A)} + \ldots.
}

As showed by \cite{Loganayagam:2011mu}, in presence of just $U(1)$ anomaly, 
it gives the exact result of Son-Sorowka \cite{Son:2009tf}. To write a similar expression for gravitational anomaly in Landau 
Frame, one will need to find anomalous velocity to subsubleading order, which might be non-trivial.

\section{Counting of Independent Terms} \label{sec:counting}

This section is dedicated to develop a systematic procedure to compute
independent fluid data (vectors, tensors transverse to velocity and
scalars). First we will review the counting in parity-even sector in
generic dimensions. Then we will
extend this idea to parity-odd sector in generic dimensions at
arbitrary derivative order through a procedure we call {\it`derivative
counting'}.

After describing the generic procedure, we explicitly construct
leading and sub-leading order parity-odd and even terms which are
important for our current work. Many of these terms vanish in
equilibrium. In tables \cref{tab:subleadingparityevendata,tab:leadingparityevendata,tab:parity2data} we list all the leading and sub-leading
terms both parity-odd and even and check if they survive at
equilibrium. Further in \cref{apn:subsubleading} we extend this counting procedure to
parity-odd subsubleading derivative order fluid. For this reason we will keep our illustrations in the 
construction explicit to subsubleading order.

\subsection{Parity-even Counting}

In this subsection we present the parity-even counting in generic
dimensions. One can always count
independent data in the local rest frame (LRF) of the fluid, which
turns out to be easier. We can later covariantize the terms to a
generic reference frame by following simple (and generic) rules\footnote{
The rules can be summarized as: replace 1) `$0$' index with contraction with $u^\mu$, 2) `$i$' indices with a projection along $P^{\mu\nu}$, 3) $\dow$ with $\hat\N$, 4) $\e^{ki_2j_2\ldots i_nj_n}$ with $\e^{\mu_1\nu_1\mu_2\nu_2\ldots \mu_n\nu_n}u_{\mu_1}$, and finally 5) put all extra factors of projectors and velocities on left-most, so no derivatives act on them.
}. In LRF, the fundamental quantities are
\begin{itemize}
	\item Temperature -- $\vq$, Chemical Potential -- $\nu$.
	\item Derivatives of fluid velocity\footnote{$u^\mu u_\mu =
            -1$ would imply $u^\mu\dow u_\mu = 0$ and hence in local
            rest frame $\dow u_0 = 0$.} -- $\dow_0 u^i$, $\dow^j u^i$.
	\item Field Tensor --  $\cF^{ij}$, $\cE^{i} = \cF^{i\nu}u_\nu$.
	\item Curvature -- $\cR^{ijkl}$, $\cR^{ijk0}$, $\cR^{i0k0}$.
\end{itemize}
All other quantities are merely
derivatives of these fundamental quantities. Since LRF is locally
flat, we are using the coordinate derivatives $\dow_o$ and
$\dow_i$. We introduce a notation for parity-even terms which will be useful
later in parity-odd counting. Terms with $d$ derivatives and $i$
indices will be denoted collectively as $(\frac{i}{d},i,d)$. When
working at equilibrium, it is also convenient to define\footnote{ Our
  conventions are: \bee{\nn \cA^{[\mu\nu]} = \half P^\mu_{\
      \a}P^\nu_{\ \b} \lb \cA^{\a\b} - \cA^{\b\a} \rb, \qquad
    \cA^{(\mu\nu)} = \half P^\mu_{\ \a}P^\nu_{\ \b} \lb \cA^{\a\b} +
    \cA^{\b\a} \rb, \qquad \cA^{\<\mu\nu\>} = \cA^{(\mu\nu)} -
    \frac{P^{\mu\nu}}{d-1} P_{\a\b}\cA^{\a\b}, } \bee{\nn A^{[ij]} =
    \half \lb A^{ij} - A^{ji} \rb, \qquad A^{(ij)} = \half \lb A^{ij}
    + A^{ji} \rb, \qquad A^{\<ij\>} = A^{(ij)} - \frac{g^{ij}}{d-1}
    g_{ij}A^{ij}.  } }: 
\bee{ 
	S^{\mu\nu} = 2\N^{(\mu}u^{\nu)}, \qquad
  	U^{\mu\nu} = 2\N^{[\mu}u^{\nu]}, 
} 
\bee{\label{E:defnX}
  	\cX_\L^{\mu\nu} = \lbr - \vq U^{\mu\nu}, P^{\mu\a}P^{\nu\b}
  	\cF_{\a\b} + \vq\nu U^{\mu\nu} \rbr , \quad \L={1,2}.
}

The purpose of above notation is revealed in Kaluza Klein formalism:
at equilibrium only spatial components of $\cX^{\mu\nu}_\L$ survive
which land exactly to $f^{ij}_\L$ defined by: 
\bee{ a_\L = \lbr
  \tilde\vq a^i, A^i \rbr, \qquad f_{\L}^{ij} = \N^i a_\L^j - \N^j
  a_\L^i.  }

In the same spirit we define 
\bee{ \cK^{\mu\nu\r\s} =
  P^{\mu\a}P^{\nu\b}P^{\r\g}P^{\s\d}\cR_{\a\b\g\d} - \lb U^{\mu\nu}
  U^{\r\s} + \half U^{\mu\r}U^{\nu\s} - \half U^{\nu\r}U^{\mu\s} \rb.
}
Only spatial components of $\cK^{\mu\nu\r\s}$ survive at equilibrium,
and they exactly match $R^{ijkl}$.  The usage of index `$\L$' is
purely to facilitate counting and computations. Similarly we define
$\vq_\L = \lbr \vq, \nu \rbr$.

{\bf \underline{Bianchi identity}:}

In counting, we will extensively use the Bianchi identity to get rid
of many terms, so it would be worth to spend some time on it. The
Bianchi Identities for Field Tensor, Vorticity and Riemann Tensor take
the form:
\bee{ \hat\N_{[\mu}\cF_{\nu\r]} = \hat\N_{[\mu}\hat\N_{\nu}u_{\r]} =
  \hat\N_{[\mu}\cR_{\nu\r]\s\d} = \cR_{[\mu\nu\r]\s} = 0.  }
However our redefined variables $\cX_\L$ and $\cK$ do not satisfy
Bianchi Identities. But nevertheless we can always use these
identities to relate
\bee{ \hat\N_{[\mu}\cX_{\L\nu\r]}, \qquad
  \hat\N_{[\mu}\cK_{\nu\r]\s\d}, \qquad \cK_{[\mu\nu\r]\s}, }
to other terms, and hence we can safely get rid of these in the
following computation. In rest frame especially (or at equilibrium in
any generic frame), one can check that $\cX_\L$ and $\cK$ also satisfy
Bianchi Identities.

\noindent
{\bf \underline{Killing equation}:}

If the theory has a unit Killing direction $\o^\mu$ we have the
following Killing equation for a general tensor
\bee{ \lie_{\omega} T^{\alpha_1\alpha_2\cdots}=0 \implies \o^\mu
  \hat\N_\mu T^{\a_1\a_2\ldots} = \sum_{k}T^{\a_1 \ldots \a_{k-1} \s
    \a_{k+1}\ldots} \hat\N_\s \o^{\a_k}, }
which in local rest frame becomes
 \bee{ \dow_0 T^{\a\b\g\ldots} = 0.
} 
Therefore if we are considering a theory at equilibrium, we do not
have to consider the $\dow_0$ derivatives.
Secondly, the Killing equation for metric $G^{\mu\nu}$ is given by
\bee{
	\hat\N^\b \o^\a + \hat\N^\a \o^\b = 0.
}
Taking $\frac{\o^\mu}{\sqrt{-\o^2}} =u^\mu$ and using Killing Equation for scalars this translates to:
\bee{
	\hat\N^\b u^\a + \hat\N^\a u^\b =  0.
}
Hence in local rest frame $S^{ij} = \dow^i u^j + \dow^j u^i = 0$.

\subsubsection{First Derivative Order}

Below, we compute all possible terms at first derivative order in LRF.
\begin{enumerate}
	\item $(2,2,1)$: $S^{ij}$, $\cX^{ij}_\L$
	\item $(1,1,1)$: $\dow^i \vq_\L$, $\boxed{\dow_0 u^i}$, $\cE^{i}$
	\item $(0,0,1)$: $S^{k}_{\ k}$, $\boxed{\dow_0 \vq_\L}$
\end{enumerate}

However all these first derivative terms are not independent
on-shell. Using first order equations of motion one can eliminate some
of them. The equations of motion are given by \cref{E:covariantCons}
(at equilibrium)
\begin{enumerate}
\item $(1,1,1)$: $\dow_\mu \bar\cT^{\mu i} = \cF^{i\a}\bar\cJ_\a + \undertilde{\fT}^{\nu}$
\item $(0,0,1)$: $\dow_\mu \bar\cT^{\mu}_{\ \ 0} = 
- \cE^{\a}\bar\cJ_\a + \undertilde{\fT}^{\nu} u_\nu$, \quad $\dow_\mu \bar\cJ^\mu = \undertilde{\fJ}$.
\end{enumerate}
Using these equations we have killed the $\boxed{\text{boxed}}$ terms
in the counting.

\subsubsection{Second Derivative Order}

Below we list all possible $pure$ second derivative terms. By pure we
mean they are not product of two first derivative terms. Product of
two lower derivative terms are called $composite$ terms.
\begin{enumerate}
	\item $(2,4,2)$: $\cK^{ijkl}$
	\item $(\frac{3}{2},3,2)$: $\dow^i S^{jk}$,
          $\dow^i\cX^{jk}_\L$, $\cR^{ijk}_{\ \ \ 0}$
	\item $(1,2,2)$: $\dow^i\dow^j \vq_\L$, $\dow^i \cE^{j}$,
          $\boxed{\dow_0 S^{ij}}$, $\boxed{\dow_0 \cX^{ij}_1}$,
          $\dow_0 \cX^{ij}_2$, $\cR^{i\ k}_{\ 0 \ 0}$, $\cK^{iaj}_{\ \
            \ a}$
	\item $(\half,1,2)$: 
          $\boxed{\dow^i \dow_0 \vq_\L}$, $\boxed{\dow_0\dow_0 u^i}$,
          $\dow_0 \cE^{i}$, $\dow_i S^{ij}$, $\dow_i\cX^{ij}_\L$,
          $\cR^{ia}_{\ \ a0}$
	\item $(0,0,2)$: $\boxed{\dow_0 S^{k}_{\ k}}$,
          $\boxed{\dow_0\dow_0\vq_\L}$, $\dow_i\dow^i \vq_\L$, $\dow_i
          \cE^{i}$, $\cK^{ab}_{\ \ ab}$, $\cR^{a}_{\ 0a0}$
\end{enumerate}

Here also all the terms are not independent because of equations of
motion. The second order equations of motion are given by,
\begin{enumerate}
\item $(1,2,2)$: $\dow^{(k} \dow_\mu \cT^{\mu i)} =
  \dow^{(k}\lb\cF^{i)\a}\cJ_\a + \undertilde{\fT}^{i)} \rb$, \quad $\dow^{[k} \dow_\mu \cT^{\mu
    i]} = \dow^{[k}\lb\cF^{i]\a}\cJ_\a + \undertilde{\fT}^{i]}\rb$
\item $(\half,1,2)$: $\dow_0\dow_\mu \cT^{\mu i} =
  \dow_0\lb\cF^{i\a}\cJ_\a + \undertilde{\fT}^{i}\rb$, \quad $\dow^i\dow_\mu \cT^{\mu}_{\ \
    0} = - \dow^i\lb\cE^{\a}\cJ_\a - \undertilde{\fT}^{\a}u_\a\rb$, \\ \ \quad $\dow^i\dow_\mu \cJ^\mu
  = \dow^i \undertilde{\fJ}$
\item $(0,0,2)$: $\dow_i\dow_\mu \cT^{\mu i} =
  \dow_i\lb\cF^{i\a}\cJ_\a + \undertilde{\fT}^{i)}\rb$, \quad $\dow_0\dow_\mu \cT^{\mu}_{\ \
    0} = - \dow_0\lb\cE^{\a}\cJ_\a - \undertilde{\fT}^{\a}u_\a\rb$, \\ \ \quad $\dow_0\dow_\mu \cJ^\mu
  = \dow_0 \undertilde{\fJ}$
\end{enumerate}

Again we have killed $\boxed{\text{boxed}}$ terms in the counting
using equations of motion. We have provided a list of all terms till
second order (also composites) in covariant form and their equilibrium values in
\cref{tab:leadingparityevendata,tab:subleadingparityevendata}. We can iterate this
procedure to further derivative orders as required by the cause. Note
that, for a pure term at $N$th derivative order, the maximum number of
indices possible are $N+2$; we will need it later.

\mktbl{t}{tab:leadingparityevendata}{Independent Leading Order
  Parity-even Data}{} {|c|c|c|c|} { \hline
  Name	& LRF	& Covariant & Equilibrium \\
  \hline\hline

	$\Q$ & $\half S^{i}_{\ i}$	& $\half S^{\mu}_{\ \mu}$	& 0 \\
	\hline\hline	
	
  $V_\L^\mu$ & $\dow^i\vq_\L$ & $P^{\mu\a}\hat\N_\a\vq_\L$ &
  $\N^i\vq_{\L o}$ \\

	\hline
	
	$V_3^\mu$	& $\cE^{i} - \vq_1 V^{i}_{2}$	& $\cE^{\mu} - \vq_1 V^\mu_{2}$ & 0 \\
	\hline\hline
	
	$\s^{\mu\nu}$ & $\half S^{\langle ij\rangle}$	& $\half S^{\langle\mu\nu\rangle}$ & 0 \\  
  
  \hline
}
\mktbl{t}{tab:subleadingparityevendata}{Independent Subleading Order
  Parity-even Data}{} {|c|c|c|c|} { \hline
  Name	& LRF	& Covariant & Equilibrium \\
  \hline\hline
	
  $\mathbf S_{1\L}$ & $\dow^i\dow_i\vq_\L$ &
  $P^{\a\b}\hat\N_\a\hat\N_\b\vq_\L$ & $\N^i\N_i\vq_{\L o}$
  \\
  \hline
	
  $\mathbf S_{2(\L\G)}$ & $\dow^i\vq_\L\dow_i\vq_\G$ &
  $P^{\a\b}\hat\N_\a\vq_\L\hat\N_\b\vq_\G$ & $\N^i\vq_{\L
    o}\N_i\vq_{\G o}$ \\
  \hline
	
  $\mathbf S_{3(\L\G)}$ & $\cX_{\L}^{ij}\cX_{\G ij}$ &
  $\cX_{\L}^{\mu\nu}\cX_{\G\mu\nu}$	& $f_{\L}^{ij}f_{\G ij}$ \\
  \hline
	
  $\mathbf S_{4}$	& $\cK$	& $\cK$	& $R$ \\
  
	\hline
	
	$\bS_5$	& $\dow_i V^{i}_{3}$	& $P_{\mu\nu}\hat\N^\mu V^\nu_{3}$ & 0 \\
	\hline
	
	$\bS_6$	& \begin{minipage}{130px}
		\centering
		$\cR_{00} 
		+ \frac{1}{\vq}\dow_i \dow^i \vq$ \\
		$ - 2\frac{1}{\vq^2}\dow_i \vq\dow^i \vq
		- \frac{1}{4}\frac{1}{\vq^2} \cX^{ij}_1 \cX_{1ij}$	
	\end{minipage}& \begin{minipage}{130px}
		\centering
		$ u^\mu u^\nu \cR_{\mu\nu} 
		+ \frac{1}{\vq} \bS_{1,1}$
		$ - 2\frac{1}{\vq^2} \bS_{2,11}
		- \frac{1}{4}\frac{1}{\vq^2} \bS_{3,11}$	
	\end{minipage} & 0 \\
	\hline
	
	$\bS_{7\L}$	& $V_\L^i V_{3i}$	& $V_\L^\mu V_{3\mu}$ & 0 \\
	\hline
	
	$\bS_8$	& $V_3^i V_{3i}$	& $V_3^\mu V_{3\mu}$ & 0 \\
	\hline
	
	$\bS_9$	& $\Q^2$	& $\Q^2$ & 0 \\
	\hline
	
	$\bS_{10}$	& $S^{ij} S_{ij}$	& $S^{\mu\nu} S_{\mu\nu}$ & 0 \\  
  
  \hline\hline
	
  $\mathbf V_{1\L}^\mu$ & $\dow_k \cX^{ki}_\L$ &
  $P^{\mu\g}P_{\a\b}\hat\N^\a \cX_{\L\g}^{\b}$ & $\N_k
  f^{ki}_\L$ \\
  \hline
	
  $\mathbf V_{2\L\G}^\mu$ & $ \cX^{ik}_\L V_{\G k}$ & $
  \cX^{\mu\a}_\L V_{\G\a} $	& $ f^{ik}_\L \N_k \vq_{\G o}$ \\
  
	\hline
	
	$\bV_3^\mu$	& $\dow_0 V_3^i$	& $P^{\mu\b}u^\a\hat\N_\a V_{3\b}$ & 0  \\
	\hline
	
	$\bV_4^\mu$	& $\dow_i S^{ij}$	& $P^{\mu\b}\hat\N^\a S_{\a\b}$ & 0  \\
	\hline
	
	$\bV_5^\mu$	& 
		$ \cR^{i}_{\ 0}
		- \frac{1}{2\vq}\dow_k \cX^{ki}_1
		- \frac{3}{2\vq^2} \cX^{ik}_1\dow_k\vq$ &  
		$ P^{\mu\b}u^\a \cR_{\a\b}
		- \frac{1}{2\vq} \bV^\mu_{1,1}
		- \frac{3}{2\vq^2} \bV^i_{2,11}$ & 0  \\
	\hline
	
	$\bV_{6\L}^\mu$	& $\Q V_\L^i$	& $\Q V_\L^\mu$ & 0  \\
	\hline
	
	$\bV_7^\mu$	& $\Q V_3^i$	& $\Q V_3^\mu$ & 0  \\
	\hline
	
	$\bV_{8\L}^\mu$	& $S^{ij} V_{\L j}$	& $S^{\mu\nu} V_{\L\nu}$ & 0  \\
	\hline
	
	$\bV_9^\mu$	& $S^{ij} V_{3 j}$	& $S^{\mu\nu} V_{3\nu}$ & 0  \\
	\hline
	
	$\bV_{10\L}^\mu$	& $\cX^{ij}_\L V_{3 j}$	& $\cX^{\mu\nu}_\L V_{3\nu}$ & 0  \\  
  
  \hline\hline
	
  $\mathbf T_{1\L}^{\mu\nu}$ & $\dow^{\langle i}\dow^{j\rangle}\vq_\L$
  & $P^{\langle\mu\a}P^{\nu\rangle\b} \hat\N_\a \hat\N_\b\vq_\L$ &
  $\N^{\langle i}\N^{j\rangle}\vq_{\L
    o}$ \\
  \hline
	
  $\mathbf T_{2(\L\G)}^{\mu\nu}$ & $\dow^{\langle
    i}\vq_\L\dow^{j\rangle}\vq_\G$ &
  $P^{\langle\mu\a}P^{\nu\rangle\b}\hat\N_\a\vq_\L\hat\N_\b\vq_\G$
  & $\N^{\langle i}\vq_{\L o}\N^{j\rangle}\vq_{\G o}$ \\
  \hline
	
  $\mathbf T_{3(\L\G)}^{\mu\nu}$ & $\cX^{\langle
    ik}_{\L}\cX^{j\rangle}_{\G k}$ &
  $\cX^{\langle\mu\a}_{\L}\cX^{\nu\rangle}_{\G \a}$ &
  $f^{\langle ik}_{\L}f^{j\rangle}_{\G k}$ \\
  \hline
	
  $\mathbf T_{4}^{\mu\nu}$ & $\cK^{\< ij \>}$ &
  $\cK^{\<\mu\nu\>}$	& $R^{\< ij\>}$ \\
  
	\hline
	
	$\bT^{\mu\nu}_5$	& $\dow^{\langle i} V^{j\rangle}_{3}$	& $\hat\N^{\langle \mu} V^{\nu\rangle}_{3}$ & 0 \\
	\hline
	
	$\bT^{\mu\nu}_6$	& \begin{minipage}{140px}
		\centering
		$\cR^{\<i \ j\>}_{\ \ 0 \ 0} 
		+ \frac{1}{\vq} \dow^{\langle i} \dow^{j \rangle}\vq$ \\
		$- 2\frac{1}{\vq^2} \dow^{\langle i}\vq\dow^{j\rangle}\vq
		- \frac{1}{4\vq^2} \cX^{\langle i}_{1 \ a}\cX^{j \rangle a}_1$	
	\end{minipage} & \begin{minipage}{140px}
		\centering
		$P^{\langle\mu\r} P^{\nu\rangle\s} u^\a u^\b \cR_{\r\a\s\b} 
		+ \frac{1}{\vq} \bT^{\mu\nu}_{1,1}$ \\
		$- 2\frac{1}{\vq^2} \bT^{\mu\nu}_{2,11}
		- \frac{1}{4\vq^2} \bT^{\mu\nu}_{3,11}$	
	\end{minipage} & 0 \\
	\hline
	
	$\bT^{\mu\nu}_{7\L}$	& $V_\L^{\langle i} V_3^{j\rangle}$	& $V_\L^{\langle \mu} V_3^{\nu\rangle}$ & 0 \\
	\hline
	
	$\bT^{\mu\nu}_8$	& $V_3^{\langle i} V_3^{j\rangle}$	& $V_3^{\langle \mu} V_3^{\nu\rangle}$ & 0 \\
	\hline
	
	$\bT^{\mu\nu}_9$	& $\Q \s^{ij}$	& $\Q \s^{\mu\nu}$ & 0 \\
	\hline
	
	$\bT^{\mu\nu}_{10}$	& $S^{\langle ik} S^{j\rangle}_{\ k}$	& $S^{\langle \mu\a} S^{\nu\rangle}_{\ \a}$ & 0 \\
	\hline
	
	$\bT^{\mu\nu}_{11\L}$	& $S^{\langle ik} \cX^{j\rangle}_{\L \ k}$	& $S^{\langle \mu\a} \cX^{\nu\rangle}_{\L \ \a}$ & 0 \\

  \hline }

\subsection{Parity-odd Counting} \label{sec:independence}

In this section we shall compute the parity-odd leading and
sub-leading derivative fluid data. Calculation in parity-odd sector is
a lot more cumbersome, even in LRF. We introduce here a scheme called
`derivative counting' to compute these terms step by step. Any
parity-odd term in $(2n)$-dimension must have a $(2n-1)$-dim
Levi-Civita involved in LRF
 \bee{ \e^{ii_2j_2\ldots i_n j_n}.  } 
 We are interested in constructing all possible scalars, vectors and
 symmetric tensors using it. A bit of thinking will reveal that one
 needs at least $(2n-2)$-rank parity-even tensors to be combined with
 $\e^{ii_2j_2\ldots i_n j_n}$ for this purpose. One can subsequently
 form a list of \emph{parity-odd data types}:
\begin{enumerate}
\item $\mathbf V_\e$: Vectors with free index on $\e$ ($2n-2$ rank
  parity-even tensor contracted with $\e$).
\item $\mathbf S$: Scalars with all indices contracted with $\e$
  ($2n-1$ rank parity-even tensor contracted with $\e$).
\item $\mathbf T_\e$: Tensors with one free index on $\e$ ($2n-1$ rank
  parity-even tensor contracted with $\e$).
\item $\mathbf V_f$: Vectors with free index not on $\e$ ($2n$ rank
  parity-even tensor contracted with $\e$).
\item $\mathbf {V}^{C}_\e$: Vectors formed of contraction of two
  non-$\e$ indices with free index on $\e$ ($2n$ rank parity-even
  tensor contracted with $\e$).
\item $\mathbf T_f$: Tensors with no free index on $\e$ ($2n+1$ rank
  parity-even tensor contracted with $\e$).
\item $\mathbf {S}^{C}$: Scalars formed of contraction of $\mathbf
  T_f$ ($2n+1$ rank parity-even tensor contracted with $\e$).
\item $\mathbf {T}^{C}_\e$: Tensors formed of contraction of two
  non-$\e$ indices with one free index on $\e$ ($2n+1$ rank
  parity-even tensor contracted with $\e$).
\item $\mathbf {V}^{C}_f$: Vectors formed of contraction of two
  non-$\e$ indices with one free index not on $\e$ ($2n+2$ rank
  parity-even tensor contracted with $\e$).
\item $\mathbf {V}^{CC}_\e$: Vectors formed of contraction of four
  non-$\e$ indices with free index on $\e$ ($2n+2$ rank parity-even
  tensor contracted with $\e$)\\
.\\
.\\
.\\
and so on.
\end{enumerate}

Here we note that given $D$ derivatives, one cannot construct a
parity-even term, pure or composite, with more than $2D$
indices, because $(2,2,1)$ and $(2,4,2)$ have the highest index to
derivative ratio, which is 2. Therefore, if we are interested in a fluid
at $(n-2 + s)$ derivative order ($s=1$ corresponds to parity-odd
leading order and so on), we can get at most $2(n-2+s)$ indices. The
list of parity-odd data types we gave above is complete till
subsubleading derivative order $(s=3)$.

\subsubsection*{Independent Data Types}

We should emphasise that not all parity-odd data-types listed above
are independent. The dependence comes from that fact that when we are
working in $2n-1$ dimensions, any antisymmetrization over $2n$ or more
indices will vanish. Given that we are dealing with parity-even
tensors of arbitrary rank which are to be contracted with $\e$, there
are a whole lot of these antisymmetrizations possible. Hence, to find
the independent data-types becomes highly non-trivial.

Let's look at a special case of this dependence. We construct a
$2n$-antisymmetrization, 
\bee{\label{E:2nassym} \e^{[i_1\ldots
    i_{2n-1}}A^{k_1]k_2\ldots k_t}_{\hspace{32pt}i_1\ldots i_{2n-1}} =
  0, }
therefore, 
\bee{\label{E:dependence} \e^{i_1\ldots
    i_{2n-1}}A^{k_1k_2\ldots k_t}_{\hspace{30pt}i_1\ldots i_{2n-1}} =
  \sum_{a=1}^{2n-1} (-1)^{a+1} \e^{k_1i_1\ldots i_{2n-2}} A^{x
    k_2\ldots k_t}_{\hspace{26pt} i_1\ldots i_{a-1} x i_a \ldots
    i_{2n-2}}.  }

The consequence of this is that the data types $[~]_f$ (i.e. ones with
a free index not on $\e$) can be expressed in terms of $[~]_\e^C$
(i.e. the ones with a free index on $\e$ and an extra
contraction). Hence data-types $[~]_f$ for example
$\bV_f,\bT_f,\bV_f^C$ are not independent.

Note that this result is only based on a specific form of
$2n$-antisymmetrization (\cref{E:2nassym}). One can in principle go on
with any random antisymmetrizations over $2n$ or more indices and find
relations among the data, which as it turns out, is not a trivial task
to do. We will come back to this issue in \cref{sec:basisdata}. For
now we continue with the counting.

\subsubsection{Derivative Counting}

We have classified parity-odd terms in data-types based on the number
of parity-even indices required. We want to construct all allowed parity odd terms with
$D$ derivatives. We observe that it is not required to include all parity-even data type 
of the form $(r,i,d)$ in this construction. We will show this below.

  For a parity-odd fluid at
$D=(n-2+s)$ derivative order, we need to construct all the $D$
derivative parity-even terms with number of indices ranging from $2D$
(the maximum possible) to $2(D+1-s)$ (= $2n-2$, the minimum
required), i.e.
$$
2(D+1-s)\leq \text{No of indices of a parity-even D derivative
  term}\leq 2D.
$$ 
These $D$-derivative parity-even terms can be constructed out of pure
derivative terms. We need not consider pure terms with self
contractions in parity-even data types as they have been included in
our counting procedure.

We now want to argue that not all parity-even data-types are required for this construction. For a data-type $(\frac i N, i, N)$ to be included at least once, the following combination with $(2D-2N+i)$ indices must be included:
\bee{
	(D-N) \times (2,2,1) \otimes (\frac i N, i,N) \nn
}
Since the minimum rank of this term
must be $2n-2 =2(D+1-s)$ and maximum possible rank is $N+2$, therefore we get,
 \bee{ N+2 \geq i\geq
  2(N-s+1).  }
For this equation to have a solution $N\leq 2s$. So we need at max $2s$ derivative order parity-even terms, to
construct parity-odd terms till $(n-2+s)$ derivative order. For example at leading order, $s=1$, only pure terms with at max $2$ derivatives are required. The parity
even terms required till $s=3$ are enlisted in
\cref{tab:classdata,tab:classdata_noneqb}. Further, if we were only interested in finding terms that survive at
equilibrium, we can use the Killing condition and drop all terms with
$\dow_0$ derivatives.

\mktbl{t}{tab:classdata}{Parity-even Data-types -- Surviving at Equilibrium}{}
{|c|c|c|c|}
{
	\hline
	Data Type	& Decomposition	& Local Rest Frame	& Equilibrium \\
	\hline
	\hline
	$(2,2,1)$	&& $\cX_\L^{ij}$	& $f_\L^{ij}$ \\
	\hline
	$(1,1,1)$	&& $\dow^i\vq_\L$	& $\N^i\vq_{\L o}$ \\
	\hline\hline
	
	$(2,4,2)$	&& $\cK^{ijkl}$ & $R^{ijkl}$ \\
	\hline
	$(\frac{3}{2},3,2)$	& $\dow^i (2,2,1)$	& $\dow^i \cX_\L^{jk}$	& $\N^i f_\L^{jk}$ \\
	\hline
	$(1,2,2)$	& $\dow^i (1,1,1)$ & $\dow^i\dow^j\vq_\L$	& $\N^i\N^j\vq_{\L o}$ \\
	\hline\hline
	
	$(\frac{5}{3},5,3)$	& $\dow^i (2,4,2)$	& $\dow^i\cK^{jklm}$ 		& $\N^i R^{jklm}$ \\
	\hline
	$(\frac{4}{3},4,3)$	& $\dow^i\dow^j (2,2,1)$ & $\dow^i\dow^j \cX_\L^{kl}$	& $\N^i\N^j f_\L^{kl}$ \\
	\hline
	$(1,3,3)$	& $\dow^i\dow^j (1,1,1)$ & $\dow^i\dow^j\dow^k\vq_\L$	& $\N^i\N^j\N^k\vq_{\L o}$ \\
	\hline\hline
	
	$(\frac{3}{2},6,4)$	& $\dow^i\dow^j (2,4,2)$	& $\dow^i\dow^j\cK^{klmn}$ 		& $\N^i\N^j R^{klmn}$ \\
	\hline
	$(\frac{5}{4},5,4)$	& $\dow^i\dow^j\dow^k (2,2,1)$ & $\dow^i\dow^j\dow^k \cX_\L^{lm}$	& $\N^i\N^j\N^k f_\L^{lm}$ \\
	\hline
	$(1,4,4)$	& $\dow^i\dow^j\dow^k (1,1,1)$ & $\dow^i\dow^j\dow^k\dow^l\vq_\L$	& $\N^i\N^j\N^k\N^l\vq_{\L o}$ \\
	\hline\hline
	
	$(\frac{7}{5},7,5)$	& $\dow^i\dow^j\dow^k (2,4,2)$	& $\dow^i\dow^j\dow^k\cK^{lmno}$ 		& $\N^i\N^j\N^k R^{lmno}$ \\
	\hline
	$(\frac{6}{5},6,5)$	& $\dow^i\dow^j\dow^k\dow^l (2,2,1)$ & $\dow^i\dow^j\dow^k\dow^l \cX_\L^{mn}$	& $\N^i\N^j\N^k\N^l f_\L^{mn}$ \\
	\hline\hline
	
	$(\frac{4}{3},8,6)$	& $\dow^i\dow^j\dow^k\dow^l (2,4,2)$	& $\dow^i\dow^j\dow^k\dow^l\cK^{mnop}$ 		& $\N^i\N^j\N^k\N^l R^{mnop}$ \\
	\hline
}

\mktbl{h}{tab:classdata_noneqb}{Parity-even Data-types -- Vanishing at Equilibrium}{}
{|c|c|c|}
{
	\hline
	Data Type	& Decomposition	 & Local Rest Frame \\
	\hline
	\hline
	$(2,2,1)$	& \multicolumn{2}{c|}{$\s^{ij} := \half S^{\<ij\>}$} \\
	\hline
	$(1,1,1)$	& \multicolumn{2}{c|}{$V_3^i := \cE^i - \vq_1 V_2^i$} \\
	\hline
	\hline
	\multirow{2}{*}{$(\frac{3}{2},3,2)$}	& \multicolumn{2}{c|}{ $\X^{ijk} := \cR^{ijk}_{\ \ \ 0}
	- \frac{1}{2\vq_o} \N^k \cX_1^{ij}
	+ \frac{1}{\vq_o^2} \lb
		f_1^{ij}\N^k\vq_o 
		+ \half \cX_1^{ik} \N^j \vq_o
		- \half \cX_1^{jk} \N^i \vq_o
	\rb $} \\
	\cline{2-3}
	& $\dow^i (2,2,1)$	& $\dow^i\s^{jk}$ \\
	\hline
	\multirow{3}{*}{$(1,2,2)$}	& \multicolumn{2}{c|}{ $\X^{ij} := \cR^{i \ j}_{\ 0 \ 0} 
	+ \frac{1}{\vq} \dow^{i} \dow^{j}\vq 
	- 2\frac{1}{\vq^2} \dow^{i}\vq\dow^{j}\vq 
	- \frac{1}{4\vq^2} \cX^{i}_{1 \ a}\cX^{ja}_1$} \\
	\cline{2-3}
	& $\dow_0 (2,2,1)$	& $\dow_0 \cX_\L^{jk}$, $\dow_0\s^{ij}$ \\
	\cline{2-3}
	& $\dow^i(1,1,1)$ & $\dow^i V_3^j$ \\
	\hline
	$(\half,1,2)$	& $\dow_0 (1,1,1)$ & $\dow_0\dow_0\vq_\L$, $\dow_0 V_3^i$ \\
	\hline
	\hline
	\multirow{3}{*}{$(\frac{4}{3},4,3)$}	& $\dow^i(\frac{3}{2},3,2)$ & $\dow^i\X^{jkl}$ \\
	\cline{2-3}
	& $\dow_0 (2,4,2)$	& $\dow_0\cK^{jklm}$ \\
	\cline{2-3}
	& $\dow^i\dow^j (2,2,1)$ & $\dow^i\dow^j\s^{kl}$ \\
	\hline
	\multirow{4}{*}{$(1,3,3)$}	& $\dow_0(\frac{3}{2},3,2)$ & $\dow_0\X^{ijk}$ \\
	\cline{2-3}
	& $\dow^i(1,2,2)$ & $\dow^i\X^{jk}$ \\
	\cline{2-3}
	& $\dow_0\dow^j (2,2,1)$ & $\dow_0\dow^j \cX_\L^{kl}$, $\dow_0\dow^i\s^{jk}$ \\
	\cline{2-3}
	& $\dow^i\dow^j(1,1,1)$ & $\dow^i\dow^jV_3^k$ \\
	\hline
	\multirow{3}{*}{$(\frac{2}{3},2,3)$}	& $\dow_0(1,2,2)$ & $\dow_0\X^{ij}$ \\
	\cline{2-3}
						& $\dow_0\dow_0 (2,2,1)$ & $\dow_0\dow_0 \cX_\L^{kl}$, $\dow_0\dow_0\s^{ij}$ \\
	\cline{2-3}
						& $\dow_0\dow^j (1,1,1)$ & $\dow_0\dow^j\dow^k\vq_\L$, $\dow_0\dow^iV_3^j$ \\
	\hline
	\hline
	\multirow{3}{*}{$(\frac{5}{4},5,4)$}	& $\dow^i\dow^j(\frac{3}{2},3,2)$ & $\dow^i\dow^j\X^{klm}$ \\
	\cline{2-3}
	& $\dow_0\dow^j (2,4,2)$	& $\dow_0\dow^j\cK^{klmn}$ \\
	\cline{2-3}
	& $\dow^i\dow^j\dow^k (2,2,1)$ & $\dow^i\dow^j\dow^k\s^{lm}$ \\
	\hline
	\multirow{5}{*}{$(1,4,4)$}	& $\dow_0\dow^i(\frac{3}{2},3,2)$ & $\dow_0\dow^i\X^{jkl}$ \\
	\cline{2-3}
				& $\dow^i\dow^j(1,2,2)$ & $\dow^i\dow^j\X^{kl}$ \\
	\cline{2-3}
				& $\dow_0\dow_0 (2,4,2)$	& $\dow_0\dow_0\cK^{klmn}$ \\
	\cline{2-3}
				& $\dow_0\dow^j\dow^k (2,2,1)$ & $\dow_0\dow^j\dow^k \cX_\L^{lm}$, $\dow_0\dow^j\dow^k\s^{lm}$ \\
	\cline{2-3}
				& $\dow^i\dow^j\dow^k(1,1,1)$ & $\dow^i\dow^j\dow^kV_3^l$ \\
	\hline
	\hline
	\multirow{3}{*}{$(\frac{6}{5},6,5)$}	& $\dow^i\dow^j\dow^k(\frac{3}{2},3,2)$ & $\dow^i\dow^j\dow^k\X^{lmn}$ \\
	\cline{2-3}
	& $\dow_0\dow^j\dow^k (2,4,2)$	& $\dow_0\dow^j\dow^k\cK^{lmno}$ \\
	\cline{2-3}
	& $\dow^i\dow^j\dow^k\dow^l (2,2,1)$ & $\dow^i\dow^j\dow^k\dow^l\s^{mn}$ \\
	\hline
}

Some of the \emph{combinations} constructed by this procedure using
\cref{tab:classdata} are:
\begin{enumerate}
	\item ($2D$ indices): $D(2,2,1)$
	\item
	\begin{enumerate}
		\item ($2D-1$ indices): $(D-1)(2,2,1) \oplus (1,1,1) $
	\end{enumerate}
	\item
	\begin{enumerate}
		\item ($2D-1$ indices): $(D-2)(2,2,1) \oplus (\frac{3}{2},3,2)$
		\item ($2D-2$ indices): $(D-2)(2,2,1) \oplus 2(1,1,1)$
		\item ($2D-2$ indices): $(D-2)(2,2,1) \oplus (1,2,2)$
	\end{enumerate}
\end{enumerate}
and so on... The counting can be extended arbitrarily to the
derivative order we need. In next section we will construct terms till
subleading order, and later in \cref{apn:subsubleading} we will extend
it to subsubleading order. We will suppress the usage of data-type
$(2,4,2)$ for brevity; combinations involving it can always be reached
by exchanging $(2,4,2)$ with two $(2,2,1)$'s.

\subsection{Examples of Parity-odd Counting}

\subsubsection{Leading Order (D=n-1) (s=1)} \label{sec:leadingcounting}

For $s=1$, the required indices are merely $2D = 2n-2$ ($\bV_\e$), which amounts to the only combination:
\bee{
	D(2,2,1),
}
along with the terms involving $(2,4,2)$. However in $\bV_\e$ all the free indices are contracted with Levi-Civita, which will kill any term involving $(2,4,2)$ due to Bianchi Identity. The only remaining combination is -- $(n)$ vectors
\bc
	$\nfrac{m-1}{n-m}^{i} \Big\vert_{m=1}^{n}$,
\ec

where we define,
\bea{
	\nfrac{m}{n-a-m}_{\mu_1\nu_1\ldots\mu_a\nu_a} 
	&= \frac{1}{2^{n-a}} \e_{\mu_{1}\nu_{1}\ldots\mu_{n}\nu_{n}}
	\prod_{x=a+1}^{m+a} \cX_1^{\mu_{x}\nu_{x}}
	\prod_{y=m+a+1}^{n} \cX_2^{\mu_{y}\nu_{y}}, \nn \\
	\nfrac{m}{n-a-m}_{ii_2j_2\ldots i_aj_a}
	&= \frac{1}{2^{n-a}} \e_{ii_{2}j_{2}\ldots i_{n}j_{n}}
	\prod_{x=a+1}^{m+a} f_1^{i_{x}j_{x}}
	\prod_{y=m+a+1}^{n} f_2^{i_{y}j_{y}}.
}

\subsubsection{Subleading Order (D=n, s=2) -- Surviving at Equilibrium} \label{sec:subleadingcounting}

At subleading order, index families required are: $2D = 2n$  ($\bV_\e^C$), $2D-1 =  2n-1$ ($\bT_\e$) and $2D-2=2n-2$ ($\bV_\e$). We only compute terms surviving at equilibrium because that is what we need for the current work.

\paragraph{2D Family:}

$2D$ family was already discussed in \cref{sec:leadingcounting}, but this time since two indices are free from $\e$, one $(2,4,2)$ can appear with two antisymmetric indices of $R^{ijkl}$ contracted. However we are supposed to take a contraction on remaining indices, which again due to antisymmetry vanish. Only remaining data are -- $(n-1)$ vectors:
\bc
	$\nfrac{m-1}{n-1-m}_{ijk}\cX^{ja}_1\cX^{k}_{2a} \Big\vert_{m=1}^{n-1}$.
\ec

\paragraph{2D-1 Family:}

Combinations in $(2D-1)$ family which survive at equilibrium are:
\begin{enumerate}
	\item $(D-1)(2,2,1)\oplus (1,1,1)$
	\item $(D-2)(2,2,1)\oplus (\frac{3}{2},3,2)$
	\item $(D-3)(2,2,1)\oplus (\frac{5}{3},5,3)$
\end{enumerate}
along with the combinations with $(2,4,2)$. In $\bT_\e$ only one index stays free from $\e$, hence again $(2,4,2)$ and $(\frac{5}{3},5,3)$ cannot appear. The remaining two combinations will yield:
\begin{enumerate}
	\item $(n-1)(2,2,1)\oplus (1,1,1)$: 2 possibilities -- $(6n-4)$ traceless symmetric tensors and $(2n)$ scalars
	\bc
		$\nfrac{m-1}{n-m}^{\langle i} \dow^{j\rangle}\vq_\L \Big\vert_{m=1}^{n}$,
		$\nfrac{m-1}{n-m-1}^{\langle ijk} \dow_j \vq_\L \cX_{\G k}^{\ \ l\rangle} \Big\vert_{m=1}^{n-1}$.
	\ec
	
	\textbf{Scalars:}
	\bc
		$\nfrac{m-1}{n-m}^{i} \dow_{i}\vq_\L \Big\vert_{m=1}^{n}$.
	\ec
	\item $(n-2)(2,2,1)\oplus (\frac{3}{2},3,2)$: 1 possibility -- $(2n-2)$ traceless symmetric tensors
	\bc
		$\nfrac{m-1}{n-m-1}^{(ijk} \dow^{l)} \cX_{\L jk} \Big\vert_{m=1}^{n-1}$.
	\ec
\end{enumerate}

\paragraph{2D-2 Family:}

Combinations in $(2D-2)$ family which survive at equilibrium are:
\begin{enumerate}
	\item $(D-2)(2,2,1)\oplus 2(1,1,1)$
	\item $(D-2)(2,2,1)\oplus (1,2,2)$
	\item $(D-3)(2,2,1)\oplus (\frac{3}{2},3,2) \oplus (1,1,1)$
	\item $(D-3)(2,2,1)\oplus (\frac{4}{3},4,3)$
	\item $(D-4)(2,2,1)\oplus 2(\frac{3}{2},3,2)$
	\item $(D-4)(2,2,1)\oplus (\frac{5}{3},5,3) \oplus (1,1,1)$
	\item $(D-4)(2,2,1)\oplus (\frac{3}{2},6,4)$
	\item $(D-5)(2,2,1)\oplus (\frac{3}{2},3,2) \oplus (\frac{5}{3},5,3)$
	\item $(D-6)(2,2,1)\oplus 2(\frac{5}{3},5,3)$
\end{enumerate}
Along with these, we have
 the combinations with $(2,4,2)$. 
However, $\bV_\e$ has no index free from $\e$, and hence Bianchi Identity 
will not allow $(2,4,2)$, $(\frac{5}{3},5,3)$ and $(\frac{3}{2},6,4)$. 
Further, $(1,2,2)$, $(\frac{3}{2},3,2)$ and $(\frac{4}{3},4,3)$ will vanish 
as they cannot be made completely antisymmetric. Finally only one combination will remain, yielding:
\begin{enumerate}
	\item $(n-2)(2,2,1)\oplus 2(1,1,1)$: 1 possibility -- $(n-1)$ vectors
	\bc
		$\nfrac{m-1}{n-m-1}^{ijk}\dow_j \vq_1 \dow_k \vq_2 \Big\vert_{m=1}^{n-1}$.
	\ec
\end{enumerate}

At equilibrium we have $(2n)$ scalars, $(2n-2)$ vectors and $(8n-6)$ traceless symmetric tensors. We have tabulated these data and their equilibrium values in \cref{tab:parity2data}. 

\mktbl{t}{tab:parity2data}{Independent Leading and Subleading Order Parity-odd Data at Equilibrium}{}
{|c|c|c|}
{
	\hline
	Name	& Term	& Equilibrium \\
	\hline\hline
	
	$l^\mu_m \big\vert_{m=1}^n$
	& $\nfrac{m-1}{n-m}^{\mu\nu}u_\nu$
	& $\nfrac{m-1}{n-m}^{i}$ \\
	\hline\hline\hline
	
	$\mathbf {\tilde S}_{\L m} \big\vert_{m=1}^n$
	& $l^{\mu}_m \hat\N_{\mu}\vq_\L$
	& $\nfrac{m-1}{n-m}^{i} \N_{i}\vq_{\L o}$ \\
	\hline\hline
	
	$\mathbf{\tilde V}_{1m}^\mu \big\vert_{m=1}^{n-1}$
	& $\nfrac{m-1}{n-1-m}_{\mu\nu\r\s}u^\nu\cX^{\r\a}_1\cX^{\s}_{2\a}$
	& $\nfrac{m-1}{n-1-m}_{kij} f^{ia}_1 f^{j}_{2 a}$ \\
	\hline
	$\mathbf {\tilde V}_{2m}^\mu \big\vert_{m=1}^{n-1}$
	& $\nfrac{m-1}{n-m-1}^{\mu\nu\r\s}u_\nu \hat\N_\r \vq_1 \hat\N_\s \vq_2 $
	& $\nfrac{m-1}{n-m-1}^{ijk}\N_j \vq_{1 o} \N_k \vq_{2 o}$ \\
	\hline\hline
	
	$\mathbf {\tilde T}^{\mu\nu}_{1\L m} \big\vert_{m=1}^{n}$
	& $l^{\langle\mu}_m P^{\nu\rangle\a}\hat\N_\a \vq_\L$
	& $\nfrac{m-1}{n-m}^{\langle i} \hat\N^{j\rangle}\vq_{\L o}$ \\
	\hline
	$\mathbf {\tilde T}^{\mu\nu}_{2\L m} \big\vert_{m=1}^{n-1}$
	& $\nfrac{m-1}{n-m-1}^{(\mu\nu\r\s}u_\nu \hat\N^{\a)}\cX_{\L\r\s}$
	& $\nfrac{m-1}{n-m-1}^{( ijk} \N^{l )} f_{\L jk}$ \\
	\hline
	$\mathbf {\tilde T}^{\mu\nu}_{3\L\G m} \big\vert_{m=1}^{n-1}$
	& $\nfrac{m-1}{n-m-1}^{(\mu\nu\r\s}u_\nu \hat\N_\r\vq_\L \cX_{\G\s}^{\ \ \d)}$
	& $\nfrac{m-1}{n-m-1}^{(ijk}\N_j\vq_{\L o} f_{\G k}^{\ \ l)}$ \\
	\hline
}

\subsection{The Basis of Independent Data} \label{sec:basisdata}

As we discussed in \cref{sec:independence}, the data we have enlisted
in the preceding sections is a `complete set' but not
independent. There might exist numerous relations among them through
antisymmetrizations of $2n$ or more indices. If we look back at
\cref{sec:scheme}, the need of all independent data arose to write
down the most generic form of the constitutive relations. We write the
energy-momentum tensor and charged current as a combination of all
independent tensors and vectors respectively up to some undetermined
coefficients which are called transport coefficients. We then
determine the same quantities from equilibrium partition function and
compare with the fluid results. It turns out that the transport
coefficients which destroys the positivity of entropy
current divergence are set to zero by this procedure. We call these transport coefficients
$unphysical$.  Put differently, the partition function generates only
the $physical$ transport coefficients in the constitutive relations
(\cref{E:parti_perturb}) at equilibrium.

Now if we relax the condition \emph{`independence'} while writing
fluid constitutive relations, $i.e.$, add more terms to these relations
which could have been determined in terms of others; they can be
regarded as redundant transport coefficients in our system. Since the
charge current and the energy-momentum tensor we derive from the partition
function remain unchanged, we get relations between the transport
coefficients (including the redundant coefficients) and the coefficients
appearing in partition function. 
However, we still have our answers -- the independent transport
coefficients and distinct constitutive relations.

Let us explain with an example. Suppose at some particular derivative
order, we have total $I$ number of vectors $V^{\mu}_i$. We can write charge current at this order as, $J^{\mu} = \sum_{i=1}^I a_i V_i^{\mu}$, where $a_i$'s are transport coefficients. On the other hand, suppose our partition function has $X$ number of independent coefficients $C_j$'s, and it generates a charge current $J^{\mu} = \sum_{i=1}^I c_i(C_j) V_i^{\mu}$. $c_i(C_j)$ are some functions of $C_j$'s. By comparison we will get $a_i = c_i(C_j)$. These are $I$ relations with $X$ free parameters, and thus imposes $I-X$ constraints on $a_i$.

Now let's add to our set $K$ more vectors $V^\mu_\a, \a = I+1,\ldots,I+K$ which could in principle be determined as: $V^{\mu}_\a = \sum_{i=1}^I C_{\a i}V^{\mu}_{i}$. Then we would have guessed our ansatz to be $J^{\mu} = \sum_{i=1}^{K+I} b_i V_i^{\mu}$, and by varying partition function we will get $J^{\mu} = \sum_{i=1}^{K+I} d_i(C_j) V_i^{\mu}$. $d_i(C_j)$ are some functions of $C_j$'s determined by relation $c_i = \lb d_i - \sum_{\a=I+1}^{I+K} d_\a C_{\a i} \rb$, as our partition function is still the same. By comparison we will get $b_i = d_i(C_j)$. These are $K+I$ relations with $X$ free parameters, and thus imposes $K+I-X$ constraints on $b_i$. We hence get exactly $K$ extra constraints, to kill the $K$ extra degrees of freedom we added in the system. But once we have imposed these constraints, we will only be left with $X$ independent transport coefficients.

However, note that we still need independent set of scalars that
enters the equilibrium partition function, for our arguments to make
sense. We check it here before we proceed. At leading order there are
no scalars. At subleading order the scalars do not have enough
indices for $2n$ or more antisymmetrizations, as a result all the
scalars we get are independent. At higher order however, it may not be
so easy to find out all the independent set of scalars.

Lets look at an example of such residual $2n$-antisymmetrization
conditions. In \cref{E:dependence} if we chose $B$ to be of the form
$\bS g^{ij}$, we will get: 
\bee{\label{eq:tracelessDep}
  \sum_{a=1}^{n-1} (-1)^{a+1} \e^{\langle p i_1\ldots
    i_{n-2}}A_{i_1\ldots i_{a-1} \ \ i_a\ldots i_{n-2}}^{\hspace{32px}
    q\rangle} = 0, 
} 
where $\<~\>$ denotes the traceless symmetric
part of a matrix. Hence one of these matrices of type $\bT_\e$ (after
making traceless) is not independent for a given $A$. A similar
argument is valid on other tensors like $\bT_\e^C$ using $\bS^C
g^{ij}$. But as we are treating all symmetric traceless tensors (of
type $[~]_\e$) to be independent, this should reflect in our final
constraints, and as we will see, it will. It turns out that till
subleading order, \cref{eq:tracelessDep} is the only remaining
residual constraint, and thus we can construct an independent basis;
but this issue might turn more subtle at higher derivative orders. To
illustrate the procedure we will not start with the independent basis
even for subleading order, and show that we get consistent results at
the end.

\section{Fluid Constitutive Relations} \label{sec:fluidrelations}

Having all the data we require, we are ready to find the constitutive
relations for fluid. We start with the results which are already known
in literature, $i.e.$ fluid up to leading derivative order. We revisit
the results in our notation. Later we consider charged fluid at
subleading order in \cref{sec:subleadingfluid}. We also set up the
notation and architecture for subsubleading order parity-odd fluid in
this formalism in \cref{apn:subsubleading}. However we do not compute
the constitutive relations explicitly, as we will discuss, the
calculation becomes a lot non-trivial.

\subsection{Ideal Fluid}

At zero derivative order only energy-momentum gets a transverse
contribution:
\bee{
	\Pi^{\mu\nu}_{(0)} = 	A P^{\mu\nu},
}
where $A$ is some arbitrary function of $\vq$ and $\nu$. Now comparing
\cref{E:CurrentsGeneric} with \cref{E:parti_perturb} we can write at
ideal order,
\bee{
	E_o v^{i}_o v^{j}_o + A_o g^{ij}
	= 
	g^{ij} P_o, \nn
}
\bee{
	- E_o v^i_o \sqrt{1+ v_{io} v^i_o} 
	- \frac{v^i_{o}}{\sqrt{1+ v_{io} v^i_o}}
	+ \vq_o\nu_o Q_o v^i_o
	= 
	0, \nn
}
\bee{
	E_o (1+ v_{io} v^i_o) 
	+ \frac{v_{io} v^i_{o}}{(1+ v_{io} v^i_o)}
	= 
	\e_o, \nn
}
\bee{
	Q_o v^i_o = 0, \nn
}
\bee{
	Q \sqrt{1+ v_{io} v^i_o}
	= 
	q_o.
}
The identifications will then give
\bee{
	v_o = -\E{\s}, \qquad 
	v^{i}_o = 0, \qquad 
	A =  P, \qquad 
	Q = q, \qquad
	E = \e.
}
Note that we have identified $A,Q,E$ exactly, and not just at equilibrium, as we explained in \cref{sec:scheme}. Therefore, the energy-momentum tensor and charge current for ideal fluid can be written as,
\bee{
	\cT^{\mu\nu}_{(0)}
	=
	\e u^\mu u^\nu
	+ P P^{\mu\nu}, \qquad
	\cJ^{\mu}_{(0)}
	=
	q u^\mu.
}

\subsection{Leading Order Fluid}

One can divide the constitutive relations in hydrodynamics in two
different sectors -- \emph{parity even} and \emph{parity odd}. Ideal fluid
belongs to the first sector (in $d>2$). The first non trivial
derivative corrections in parity-even sector appears at the first
derivative order \emph{e.g.} shear viscosity term in energy-momentum
tensor. Whereas in the parity-odd sector, the leading terms appear at
$(n-1)$ derivative order for a fluid in $2n$ dimensions. All these
terms and the corresponding transport coefficients (at leading order) have already been
found in \cite{Banerjee:2012cr}. We shall discuss their result in our notation.

\subsubsection{Parity-odd}
Since there is no parity odd scalar and transverse symmetric traceless
tensor at $(n-1)$ derivative order (see \cref{tab:parity2data}), only
charge current gets parity-odd corrections:
\bee{\label{E:-1cons}
	\tilde\U^\mu_{(n-1)} = \sum_{m=1}^{n} \ ^{n-1}\bC_{m-1} \o_m l^\mu_m.
}

The combinatorial factor is introduced for convenience. It also
ensures we do not surpass the limits of $m$. The fluid variables
receives following corrections,
\begin{equation}
	\vq_\L  = \vq_{\L o} + \tilde\D^{(n-1)} \vq_\L, \qquad v^i=
        v^i_o + \tilde\D^{(n-1)} v^i. 
\end{equation}

Further, there is no parity-odd gauge invariant scalar at equilibrium
on $\cM_{2n-1}$, implying that $\tilde\D^{(n-1)}W^{eqb}_{(C)}=0$. Now
comparing \cref{E:CurrentsGeneric} with \cref{E:parti_perturb} we will
find the constraints at parity-odd leading derivative order: \bee{
  \o_{m} = - \frac{\vq^2 n}{\e + P} \lB s C_{m-1} + q C_m + (n+1) \lb
  \frac{s}{m} + \frac{q\nu}{m+1} \rb C^{(2n)} \nu^{m} \rB.  }

And the corrections to fluid variables,
\bee{
	\tilde\D^{(n-1)} \vq =
	\tilde\D^{(n-1)} \nu = 0, \qquad
	\tilde\D^{(n-1)} v^i =
	\sum_{m=1}^{n} \ ^{n-1}\mathbf C_{m-1} \a_{o(m)} l^i_{o(m)},
}
where,
\bee{
	\a_m
	=
	- \frac{\vq^{2}n}{\e + P}\lB 
		C_{m-1} \nu
		- C_{m} 
		+ \frac{(n+1)}{m(m+1)} C^{(2n)} \nu^{m+1} 
	\rB.
}

 Here we
present these relations for completion as well as to set up our notations and
conventions. We would also like to make some interesting observations
about these functions. One can verify that
\bee{\label{E:gudrslt1} s \a_m + q \a_{m+1} = \nu \o_m - \o_{m+1}
  \qquad \forall \ m \in \lbr 1,n-1 \rbr, }
\bee{\label{E:gudrslt2}
	P^{(1,0)}\o_m = 
	s \lb P^{(1,0)} \a_{m} \rb^{(0,1)} 
	+ q \lb P^{(1,0)} \a_{m+1} \rb^{(0,1)}
	\qquad \forall \ m \in \lbr 1,n-1 \rbr.
}
Here pressure $P(\vartheta, \nu)$ is function of temperature
$\vartheta$ and redefined chemical potential $\nu$. For any function $Q(\vq,\nu)$ we define $Q^{(m,n)} = \frac{\dow^{m+n}}{\dow^m\vq\dow^n\nu} Q$. These will come
handy in subleading order calculation.

\subsubsection{Parity-even}

The most generic current corrections at parity-even leading derivative
order are (see \cref{tab:leadingparityevendata}):
\bee{ \U^\mu_{(1)} = \sum_{\L=1}^{3} \l_{\L} V^\mu_{\L}, \qquad
  \Pi^{\mu\nu}_{(1)} = -2\eta \s^{\mu\nu} - \z P^{\mu\nu} \Q, }
while at equilibrium the only surviving contributions are:
\bee{
	\U^\mu_{o(1)} = \sum_{\L=1}^{2} \l_{o\L} V^\mu_{o\L}, \qquad
	\Pi^{\mu\nu}_{o(1)} = 0.
}
There are no gauge-invariant parity-even scalars at equilibrium that
appear at this order. Therefore, $\D^{(n-1)}W^{eqb}_{(C)}=0$. Now
comparing \cref{E:CurrentsGeneric} with \cref{E:parti_perturb} we will
find at parity-even leading derivative order that all corrections
vanish
\bee{ \pi^{ij}_{o(1)}
  =\vs^{i}_{o(1)} = \D^{(1)}\vq = \D^{(1)}\nu = \D^{(1)} v^i = 0.  }
We hence get the constraints:
\bee{
	\l_1 = \l_2 = 0.
}

So finally the form of currents is
\bee{\label{E:+1cons}
	\U^\mu_{(1)} = \l_{3} V^\mu_{3}, \qquad
	\Pi^{\mu\nu}_{(1)} = -2\eta \s^{\mu\nu} - \z P^{\mu\nu} \Q.
}
We also get to know that no fluid quantities ($\vq, \nu, v^i$) get
order one parity-even correction.

\subsection{Subleading Order Fluid} \label{sec:subleadingfluid}

In this section, we shall describe the constraints on charged fluid in
arbitrary even dimensions at subleading derivative order ($i.e. \ n$
order), in presence of $U(1)$ anomaly. Where as, the subleading
correction to parity-even sector comes at second order in derivative
expansion. Some aspects of four dimensional fluids at sub-leading order have already been performed 
in \cite{Bhattacharyya:2013ida,Megias:2014mba}.

\subsubsection{Parity-odd}

Sub-leading order parity-odd fluid dynamics in four spacetime
dimensions has already been discussed in \cite{Bhattacharyya:2013ida}. Here, we generalize
the results in arbitrary even dimensions and find the constraints
on the transport coefficients. We see that, much like in \cite{Bhattacharyya:2013ida}, the higher dimensional
transport coefficients depend on first order transport coefficients $\eta, \zeta$.

From counting we can see that the $n$ order parity-odd corrections
(at eqb.) are given by (see \cref{tab:parity2data})
\bea{
	\tilde\U^\mu_{o(n)} &= 
	\sum_{m=1}^{n-1} \ ^{n-2}\bC_{m-1} \lb 
		\tilde\nu_{o1,m} \mathbf{\tilde V}_{o1,m}^{\mu} 
		+ \tilde\nu_{o2,m} \mathbf{\tilde V}_{o2,m}^{\mu}
	\rb, \\
	\tilde\Pi^{\mu\nu}_{o(n)} &= 
	\sum_{m=1}^{n-1} \ ^{n-2}\bC_{m-1} \lb 
		\tilde\t_{o1,\L m} \mathbf{\tilde T}_{o1,\L m}^{\mu\nu} 
		+ \tilde\t_{o2,\L\G m} \mathbf{\tilde T}_{o2,\L\G m}^{\mu\nu}
	\rb
	+ \sum_{m=1}^{n-1} \ ^{n-1}\bC_{m-1} \tilde\t_{o3,\L m}
        \mathbf{\tilde T}_{o3,\L m}^{\mu\nu} \nn \\ 
	& \qquad + P^{\mu\nu}_o \sum_{m=1}^{n} \ ^{n-1}\bC_{m-1} 
        \tilde\s_{o\L m} \mathbf{\tilde S}_{o\L m}. 
\label{subleading-expression}}
Sum over the relevant `$\L,\G$' indices is understood. We explicitly
write the $m$ index contraction to emphasize that the sum runs over
different values for different terms. We do not state non-equilibrium
contributions as they won't be required in this computation. 

From \cref{E:CurrentsGeneric} and \cref{E:parti_perturb} we get,
\be\label{E:Tbar}
\Delta^{(n)}\bar T = \E{2\sigma} \tilde \Delta^{(n)}\e = \E{2\sigma} \vq_o^2
\frac{\delta W_{(C)}^{eqb}}{\delta \vq_o},
\ee
\be\label{E:Jbar}
\Delta^{(n)}\bar J = - \E{\sigma} \tilde \Delta^{(n)} q = -  \E{\sigma}
\frac{\delta W_{(C)}^{eqb}}{\delta \nu_o}.
\ee
Now, 
\be
\tilde \Delta^{(n)}\e = \bfrac{\partial \e}{\partial \vq}_o \tilde \Delta^{(n)}\vq + \bfrac{\partial \e}{\partial \nu}_o \tilde\Delta^{(n)}\nu, 
\qquad \tilde \Delta^{(n)} q = \bfrac{\partial q}{\partial \vq}_o \tilde\Delta^{(n)}\vq + \bfrac{\partial q}{\partial \nu}_o \tilde \Delta^{(n)}\nu.
\ee
Therefore from \cref{E:Tbar} and \cref{E:Jbar} we can write,
\be
\tilde\D^{(n)} \vq_\L = 
	\vq_o \mathbf E_{o\L\G} \frac{ \d\tilde\D^{(n)}
          W_{(C)}^{eqb}}{\d \vq_{o\G}},
\ee
where,
\bee{ \bE_{\L\G} = \begin{pmatrix} \vq
    \frac{\dow\vq}{\dow\e}\big\vert_q, &
    \frac{1}{\vq}\frac{\dow\vq}{\dow q}\big\vert_\e \\
    \vq \frac{\dow\nu}{\dow \e}\big\vert_q, & \frac{1}{\vq}
    \frac{\dow\nu}{\dow q}\big\vert_\e
              \end{pmatrix}, \qquad
	\vq_\L = \lbr \vq, \nu \rbr.
}

Similarly comparing the $\bar T^i$ equations in
(\ref{E:CurrentsGeneric}) and (\ref{E:parti_perturb}) we get,
\be
\frac{\e_o + P_o}{\vq_o} \tilde \Delta^{(n)} v^i = \vq_o \nu_o \frac{\partial
  W_{(C)}^{eqb}}{\partial A_i} - \E{-\s}\frac{\partial W_{(C)}^{eqb}}{\partial a_i},
\ee
which can be written as,
\bee{\tilde\D^{(n)} v^i
	= (-)^\L \frac{\mu_{o\L}}{P_o^{(1,0)}} \frac{ \d\tilde\D^{(n)}
          W_{(C)}^{eqb}}{\d a_{\L i}}, \label{E:suboddcorrections}
}
where,
\bee{
\bA_{\L} = \dow_\G P \bE_{\G\L} = \begin{pmatrix}
		\vq \frac{\dow P}{\dow \e}\big\vert_q, &
		\frac{1}{\vq} \frac{\dow P}{\dow q}\big\vert_\e
	\end{pmatrix}, \qquad
	\mu_\L = \lbr \vq, \nu\vq \rbr, \qquad
	a^i_\L = \lbr\tilde\vq a^i, A^i \rbr.
}

One can check that $\bE_{\L\G}$ is symmetric matrix and $\dow_\L =
\frac{\dow}{\dow \vq_\L}$. We would like to emphasize that these are
purely notations, to make the calculations tractable and easy to
digest. There is a summation on repeated $\L,\G$ indices.  Now comparing $\bar T^{ij}$ and $\bar J^i$ in \cref{E:CurrentsGeneric}
with \cref{E:parti_perturb} at parity-odd subleading derivative order,
we have corrections to constitutive relations
\bea{ \frac{1}{\vq_o}\tilde\pi^{ij}_{o(n)} &= 2\frac{ \d\tilde\D^{(n)}
    W_{(C)}^{eqb}}{\d g_{ij}} - g^{ij} \mathbf A_{o\L} \frac{
    \d\tilde\D^{(n)} W_{(C)}^{eqb}}{\d \vq_{o\L}}
  - \frac{1}{\vq_o} \tilde\D^{(n-1)} \pi^{ij}_{(1)}, \nn \\
  P_o^{(1,0)}\tilde\vs_{o{(n)}}^{i} &= \vq_o \bS_{o\L} \frac{ \d
    \tilde\D^{(n)} W_{(C)}^{eqb}}{\d a_{\L i}} -
  P_o^{(1,0)}\tilde\D^{(n-1)}\vs_{(1)}^{i}, \label{E:suboddidentifications}
}
where,
\bee{
	\bS_\L = \frac{\dow P}{\dow \mu_{\bar \L}} = \lbr q, s \rbr.
}

$\bar\L$ swaps the value of $\L: 1\leftrightarrow 2$. The generating functional $\tilde\D^{(n)}W_{(C)}^{eqb}$ contain all
scalars $\mathbf{\tilde S}_{o\L m}$. But one can check that
$\mathbf{\tilde S}_{o1 m}$ can be connected to $\mathbf{\tilde S}_{o2
  m}$ by a total derivative. So we take the partition function
\bee{ \tilde\D^{(n)}W_{(C)}^{eqb} = \int \df^{2n-1} x \sqrt{g}
  \sum_{m=1}^n \ ^{n-1}\bC_{m-1} \tilde\cS_{m} \mathbf{\tilde
    S}_{o2,m}.  }
We compute the variation of generating functional with respect to
different fields and find that
\bea{
  \frac{\d \tilde\D^{(n)} W_{(C)}^{eqb}}{\d g_{ij}} &= 0, \nn \\
  \frac{\d \tilde\D^{(n)}W_{(C)}^{eqb}}{\d \vq_{o\L}} &=
  - (-)^\L \sum_{m=1}^n \ ^{n-1}\bC_{m-1} \tilde\cS_{m}^{(1,0)} \mathbf{\tilde S}_{o\bar\L,m}, \nn \\
  \frac{\d \tilde\D^{(n)}W_{(C)}^{eqb}}{\d a_{\L i}} &=
  (n-1)\sum_{m=1}^{n-1} \ ^{n-2}\bC_{m-1} \tilde\cS_{m+2-\L}^{(1,0)}
  \mathbf{\tilde V}^i_{o2,m}.  \label{acn-variation}}
Using the form of lower order currents corrections from
\cref{E:+1cons} we can write,
\bea{
	\tilde\D^{(n-1)} \pi^{ij}_{(1)} 
	&= 
	- 2\eta_o \tilde\D^{(n-1)} \s^{ij} 
	- \z_o g^{ij} \tilde\D^{(n-1)} \Q \nn \\
	&=
	- 2\eta_o\vq_o  \ ^{n-1} \bC_{m-1}\dow_\L\lb\frac{\a_{o(m)}}{\vq_o}\rb \tilde\bT^{ij}_{o1,\L m}
	- \eta_o \ ^{n-2} \bC_{m-1} (n-1) \a_{o(m+2-\L)} \mathbf{\tilde T}^{ij}_{o2,\L m} \nn \\
	& \qquad\qquad - g^{ij} \z_o \vq_o \ ^{n-1} \bC_{m-1} \dow_\L\lb\frac{\a_{o(m)}}{\vq_o}\rb \mathbf{\tilde S}_{o\L m} \\
	\tilde\D^{(n-1)} \vs^{i}_{(1)} 
	&= 
	\l_{o3} \tilde\D^{(n-1)} \cE^i \nn \\
	&=
	\l_{o3} \ ^{n-2}\bC_{m-1} (n-1) \lB
		\a_{o(m+1)}
		+ \nu_o \a_{o(m)}
	\rB
	\mathbf{\tilde V}^i_{o1m}.
\label{lowerorder-correction}}
One can now use the results, obtained in \cref{acn-variation} and
\cref{lowerorder-correction} in
\cref{E:suboddidentifications} and comparing these expressions with
\cref{subleading-expression} to get the constraints,
\bee{
	\tilde\t_{1,\L m}
	=
	2\eta \vq  \dow_\L\lb\frac{\a_{m}}{\vq}\rb, \qquad
	\tilde\t_{2,\L m}
	=
	\eta (n-1) \a_{(m+2-\L)}, \qquad
	\tilde\t_{3,\L\G m}
	=
	0,
}
\bee{
	- \frac{1}{\bA_2}\lb
		\tilde\s_{1m} 
		- \frac{\z}{2\eta} \tilde\t_{1,1 m}
	\rb
	= 
	\frac{1}{\bA_1} \lb
		\tilde\s_{2m} 
		- \frac{\z}{2\eta} \tilde\t_{1,2 m}
	\rb
	= 
	P^{(1,0)}\tilde\c_{m},
}
\bee{
	\tilde\nu_{1m} = - \l_{3} (n-1) \lb
		\a_{m+1}
		+ \nu \a_m
	\rb, \qquad
	\tilde\nu_{2m}= 
	- (n-1) \lb
		q \tilde\c_{m+1}
		+ s \tilde\c_m
	\rb.
}
Hence everything is determined in terms of a known function
$\a_m$ and a new coefficient $\tilde\c_{m}$. Note that if we had used
the $2n$-assymetrization condition \cref{eq:tracelessDep} to get rid
of one traceless symmetric tensor to start with; a consistent choice
would have been to remove $\mathbf{\tilde T}_{3,\L 1 m}^{\mu\nu}$ entirely
and $\mathbf{\tilde T}_{3,\L 2 m}^{\mu\nu}$ for $m=1$ (see
\cref{tab:parity2data}). The coefficients of these terms are set to
zero already by our constraints, which means the other leftover
constraints are independent.

Finally we get the corrections to fluid variables using
\cref{E:suboddcorrections} as
\bea{ \tilde\D^{(n)} \vq_\L &=
  \ ^{n-1}\bC_{m-1} (-)^\G \mathbf E_{o\L\G} \c_{o(m)} \mathbf{\tilde S}_{o\bar\G,m}, \nn \\
  \tilde\D^{(n)} v^i &= (n-1)\sum_{m=1}^{n-1} \ ^{n-2} \bC_{m-1} \lb
  \c_{o(m+1)} - \nu_o \c_{o(m)} \rb \mathbf{\tilde V}^i_{o2,m}.  }

\subsubsection{Parity-even}
Next, we present the results for sub-leading order (two-derivative) parity even sector for the fluid.
From counting we can verify that at the second order, parity-even corrections (at eqb.) are 
given by (see \cref{tab:subleadingparityevendata}):
\bee{
	\U_{o(2)}^\mu = \sum_\# \nu_{o\#} \mathbf V_{o\#}^\mu, \qquad
	\frac{1}{\vq_o}\Pi_{o(2)}^{\mu\nu} = \sum_\# \t_{o\#} \mathbf T_{o\#}^{\mu\nu} + P^{\mu\nu}_o\sum_\# \s_{o\#} \mathbf S_{o\#}.
}

$\#$ refers to sum over all relevant indices. Now comparing \cref{E:CurrentsGeneric} with \cref{E:parti_perturb} at parity-even subleading derivative order, and performing a similar manipulation as last section, we have corrections to constitutive relations:
\bea{
	\frac{1}{\vq_o} \pi^{ij}_{o(2)}
	&= 
	2 \frac{ \d\D^{(2)} W_{(C)}^{eqb}}{\d g_{ij}}
	- g^{ij} \bA_{o\L} \frac{ \d\tilde\D^{(2)} W_{(C)}^{eqb}}{\d \vq_{o\L}}
	- \boxed{P^{(1,0)}_o \tilde\D^{(1)} v^{\langle i} \tilde\D^{(1)} v^{j \rangle} } \nn \\
	& \qquad + \boxed{g^{ij} \tilde\D^{(1)}v_k \lbr
		\bA_{o2} \tilde\vs^{k}_{o(1)}
		+ \lb
			\frac{1}{\vq_o} \bA_{o1} P_o^{(1,0)}
			+ \frac{1}{2\vq_o} \bA_{o2} P_o^{(0,1)}
			- \frac{1}{3} P_o^{(1,0)}
		\rb \tilde\D^{(1)} v^k 
	\rbr }, \nn \\
	P^{(1,0)}_o \vs^{i}_{o(2)}
	&= 
	\vq_o \bS_{o\L} \frac{ \d\D^{(2)} W_{(C)}^{eqb}}{\d a_{\L i}}
	- \boxed{P^{(1,0)}_o \tilde\D^{(1)}\tilde\vs^{i}_{(1)}}, \label{E:subevenidentifications}
}

while the fluid variables get the corrections:
\bea{
	\tilde\D^{(2)} {\vq_\L} &= 
	\vq_o \mathbf E_{o\L\G} \frac{ \d\tilde\D^{(2)} W_{(C)}^{eqb}}{\d \vq_{o\G}}
	- \boxed{ \lb
		\bA_{o\L}
		- \half P_o^{(0,1)}\bE_{o\L 2} 
	\rb \tilde\D^{(1)} v_i \tilde\D^{(1)}v^i}
	- \boxed{\vq_o \mathbf E_{o\L 2}\tilde\D^{(1)}v_i \tilde\vs^{i}_{o(1)}}, \nn \\
	\D^{(2)}v^i 
	&= 
	(-)^\L \frac{\mu_{o\L}}{P^{(1,0)}_o} \frac{ \d\D^{(2)} W_{(C)}^{eqb}}{\d a_{\L i}}. \label{E:subevencorrections}
}

Notice that the boxed terms only contribute for four dimensional fluids ($n=2$). Out of the 
scalars enlisted in \cref{tab:subleadingparityevendata}, $\bS_{o1\L}$ can be related to others by a total 
derivative. Hence $\D^{(2)} W_{(C)}^{eqb}$ is given by:
\bee{
	\D^{(2)} W_{(C)}^{eqb}
	=  
	- \half \int \lbr\df x^i\rbr \sqrt{g}\lbr
		\cS_R \bS_{o4}
		+ \cS_{f\L\G} \bS_{o3(\L\G)}
		+ \cS_{\vq\L\G} \bS_{o2(\L\G)}
	\rbr.
}

Now we can find the variations of $\D^{(2)} W_{(C)}^{eqb}$,
\bem{
	2\frac{\d \D^{(2)} W_{(C)}^{eqb}}{\d g_{ij}} =
	- \dow_\L\cS_R \mathbf T_{o1,\L}^{ij}
	- \lb \dow_\L\dow_\G\cS_R - \cS_{\vq \L\G} \rb \mathbf T_{o2,\L\G}^{ij}
	+ 2 \cS_{f\L\G} \mathbf T_{o3,\L\G}^{ij}
	+ \cS_R \mathbf T_{o4}^{ij} \\
	+ g^{ij}\lB
		\lb
			1
			- \frac{1}{d-1} 
		\rb \dow_\L\cS_R \mathbf S_{o1,\L}
		+ \lb 
			\dow_\L\dow_\G\cS_R 
			- \frac{1}{d-1} \dow_\L\dow_\G\cS_R 
			- \half \cS_{\vq \L\G}
			+  \frac{1}{d-1} \cS_{\vq \L\G} 
		\rb \mathbf S_{o2,\L\G} \dbrk
		- \half \lb
			1
			- \frac{4}{d-1}
		\rb \cS_{f\L\G} \mathbf S_{o3,\L\G}
		- \half \lb
			1
			- \frac{2}{d-1} 
		\rb \cS_R \mathbf S_{o4}
	\rB,
}
\bee{
	\frac{\d \D^{(2)} W_{(C)}^{eqb}}{\d \vq_{o\Sigma}} =
	\cS_{\vq \Sigma\L} \mathbf S_{o1,\L}
	+ \lb
		\dow_{(\G} \cS_{\vq \L)\Sigma}
		- \half \dow_\Sigma \cS_{\vq \L\G} 
	\rb \bS_{o2,\L\G}
	- \half \dow_\Sigma\cS_{f\L\G} \mathbf S_{o3,\L\G}
	- \half \dow_\Sigma\cS_R \mathbf S_{o4},
}
\bee{
	\frac{\d \D^{(2)} W_{(C)}^{eqb}}{\d a_{\L i}} =
	2 \cS_{f\L\G} \mathbf V_{o1,\G}^{i}
	- 2 \dow_\Sigma \cS_{f\L\G} \mathbf V_{o2,\G\Sigma}^i.
}

Using the form of lower order corrections from \cref{E:-1cons} for $n=2$ we can write,
\bea{
	\tilde\D^{(1)}\tilde\vs^i_{(1)}
	& = \o_{o\Sigma} l^i_{o\Sigma} \nn \\
	& =
	(-)^\Sigma \o_{o\Sigma} \lbr
		\mu_{o\bar\Sigma} \a_{\bar\L} \bV^{i}_{o1,\L}
		- \dow_\G\lb\mu_{o\bar\Sigma} \a_{\bar\L}\rb \bV^{i}_{o2,\L\G}
	\rbr.
}

We can now put the variations of generating functional along with lower order corrections worked out 
above in \cref{E:subevenidentifications}. Using \cref{E:gudrslt1,E:gudrslt2} and eliminating partition 
function coefficients $\cS$'s we will find following 7 constraints,
\bee{
	\t_{1,\L} + \dow_\L\t_{4} = 0,
}
\bee{
	\s_{o1,\L} =
	\frac{d-2}{d-1}\dow_\L\t_{4}
	- \bA_{\Sigma} \dow_\Sigma \dow_\L \t_{4} 
	- \bA_{\Sigma} \t_{2,\Sigma\L},
}
\bee{
	2\s_{2,\L\G} =  
	\dow_\L\dow_\G\t_{4}
	- \bA_{\Sigma} \dow_\Sigma \dow_\L\dow_\G\t_{4}
	- \frac{d-3}{d-1} \t_{2,\L\G}
	- 2 \bA_{\Sigma} \dow_{(\L} \t_{2,\G)\Sigma}
	+ \bA_{\Sigma} \dow_{\Sigma} \t_{2,\L\G},
}
\bee{
	4 \s_{3,\L\G} =  
	- \frac{d-5}{d-1} \t_{3,\L\G}
	+ \bA_{\Sigma} \dow_\Sigma\t_{3,\L\G},
}
\bee{
	2 \s_{4} =
	-  \frac{d-3}{d-1}\t_{4}
	+ \bA_{\Sigma}\dow_\Sigma\t_{4},
}
\bee{
	\nu_{1,\L} = 
	\frac{\vq}{P^{(1,0)}} \bS_\G \t_{3,\L\G},
}
\bee{
	\nu_{2,\G\L} = 
	- \frac{\vq}{P^{(1,0)}} \bS_\Sigma \dow_\L \t_{3,\G\Sigma}.
}

Coincidently none of the constraints depend on $n=2$ special contributions. On the other hand fluid 
variables corrections are given by \cref{E:subevencorrections}:
\bem{
	\tilde\D^{(2)} {\vq_\O} =
	\vq_o \mathbf E_{o\O\Sigma} \lB
		\lb \t_{o2,\Sigma\L} + \dow_\L\dow_\Sigma\t_{o4} \rb \mathbf S_{o1,\L}
		+ \lb
			\dow_{(\G}\t_{o2,\L)\Sigma} 
			- \half \dow_\Sigma \t_{o2,\L\G} 
			+ \half \dow_\Sigma \dow_\L\dow_\G\t_{o4}
		\rb \bS_{o2,\L\G} \dbrk
		- \frac{1}{4} \dow_\Sigma \t_{o3,\L\G} \bS_{o3(\L\G)}
		- \half \dow_\Sigma \t_{o4} \mathbf S_{o4}
	\rB \\
	+ \boxed{ \half  \lB
		\half \vq_o \mathbf E_{o\O \Sigma} \dow_\Sigma \lb P_{o}^{(1,0)} \a_{o\bar\L}\a_{o\bar\G} \rb
		- \bA_{o\O} \a_{o\bar\L}\a_{o\bar\G}
		+ \vq_o\bE_{o\O 2} \a_{o\bar\L} \lb
			\half q_o \a_{o\bar\G}
			- \o_{o\bar\G}
		\rb
	\rB \bS_{o3(\L\G)} },
}

\bee{
	\D^{(2)}v^i 
	=
	(-)^\L \frac{\mu_{o\L}}{P^{(1,0)}_o}
	\lB
		\lb 
			\t_{o3,\L\G} 
			- \boxed{P_{o}^{(1,0)} \a_{o\bar\L}\a_{o\bar\G}} 
		\rb \mathbf V_{o1,\G}^{i}
		- \dow_\Sigma \lb
			\t_{o3,\L\G} 
			- \boxed{P_{o}^{(1,0)} \a_{o\bar\L}\a_{o\bar\G}}
		\rb \mathbf V_{o2,\G\Sigma}^i
	\rB
}

This completes our calculation of subsubleading derivative order fluid.


\section{Conclusions}

In this paper we computed the energy momentum tensor and charge current
for a fluid system in $2n$ dimensions with $U(1)$ anomaly up to
subleading order in derivative expansion (for both parity odd and
parity even sectors) from the
equilibrium partition function of the fluid. We described a novel counting prescription to construct the fluid 
data. However, an important issue we encountered here is that it is non-trivial to find independent vectors and tensors at arbitrary derivative order. But we were still able to find the independent
transport coefficients and distinct constitutive relations. We showed that the knowledge of 
independent scalars at the required derivative order is sufficient for this purpose. This is a powerful observation 
and it enables us to carry on the computation at $(n+1)$ derivative order, where, we could find the independent scalars.
We observe that the
parity odd transport coefficients which appear at $n$ derivative order
in constitutive relations are constrained and some of them depend on
the first order transport coefficients like $\eta$, $\zeta$
etc. It would be interesting to find the similar dependence in a holographic set up \cite{Banerjee:2010zd}. 
We plan to explore the holographic computation in future.

It is also interesting to find the fluid constitutive relations in presence of both $U(1)$ and gravitational anomaly in
arbitrary $2n$ dimensions.
But, since the gravitational anomaly appears at two higher derivative level compared to the $U(1)$ anomaly,
it requires to carry on our analysis to one higher derivative (sub-sub-leading) order, i.e. to $(n+1)$ derivative order. 
Fortunately, as mentioned earlier, even at this order,
we could determine the independent scalars and hence, in principle, the computation is possible. We have 
carried a large part of it in \cref{apn:subsubleading}.

\section*{Acknowledgements}

We are thankful for valuable discussions and suggestions from many of
our colleagues and friends, from which this project has been benefited
-- Felix Haehl, Sukruti Bansal, Aranya Lahiri, Pratik Roy to name a
few. A special thanks to Mukund Rangamani and R. Loganayagam for
useful discussion regarding this project. We also thank
S. Bhattacharyya for initial collaboration. AJ would also like to
thank IISER Bhopal, IISER Pune and Durham University for their
support, which made this project possible. NB would like to
acknowledge DST for Ramanujan fellowship. Finally, we are indebted to the people of India for their support.

\appendix

\section{Kaluza-Klein Decomposition} \label{apn:KK}

If a $(d+1)$-dim spacetime $\cM_{(d+1)}$ has a preferred time-like direction $\o^\mu$, it can be decomposed into $S^1 \times \cM_{(d)}$, where $S^1$ is the euclidean time circle. A $k$-rank tensor decompose in $2^k$ parts in this scheme:
\begin{enumerate}
	\item $\cS$ does not decompose.
	\item $\cV^\mu$ decompose in $\o^\mu V_\mu$ and $P^{\mu\nu} V_\nu$.
	\item $\cT^{\mu\nu}$ decompose in $\o^\mu \o^\nu T_{\mu\nu}$, $\o^\mu P_o^{\a\nu} T_{\mu\nu}$, $P_o^{\a\mu} \o^\nu T_{\mu\nu}$ and $P_o^{\a\mu} P_o^{\b\nu} T_{\mu\nu}$,
\end{enumerate}
and so on. Where $P_o^{\mu\nu} = G^{\mu\nu} - \frac{\o^\mu \o^\nu}{G_{\a\b}\o^\a \o^\b}$ is the projection operator. If we are studying theory at equilibrium, we already have a preferred direction along the Killing vector of the theory $\o^\mu = \dow_0$. In this case we know that a $(d+1)$-vector $\cV^\mu$ will yield a scalar:
\bee{
	\o^\mu \cV_\mu \Ra \cV_0 := V,
}
and a $(d)$-vector:
\bee{
	P_o^{\mu\nu}\cV_\nu \Ra \cV^i := V^i.
}
Hence we see that a $U(1)$ gauge field $\cA^\mu$ will be decomposed in $\lbr \cA_0 (\vec x), \cA^i (\vec x) \rbr$. Similarly a tensor $\cT^{\mu\nu}$ decomposes in $\cT_{00}$, $\cT^{i}_{\ 0}$, $\cT^{\ i}_0$, $\cT^{ij}$. It is the similar way the metric $G^{\mu\nu}$ on $\cM_{(d+1)}$ decomposes, hence we define:
\bee{
	G_{00} = -\E{2\s}, \quad G^{i}_{\ 0} = 0, \quad G^{ij} = g^{ij},
}
where we define $g^{ij}$ as metric on $\cM_{(d)}$. Now using the diffeomorphic invariance one can work out the full form of $G^{\mu\nu}$
\bee{
	\df s^2 
	= G_{\mu\nu}\df x^\mu \df x^\nu
	= -\E{2\s(\vec x )} \lb \df t + a_i (\vec x) \df x^i \rb^2
	+ g_{ij}(\vec x)\df x^i \df x^j,
}
\bee{
G_{\mu\nu} =
 \begin{bmatrix}
	-\E{2\s}	& -\E{2\s}a_j \\
	-\E{2\s}a_i	& \lb g_{ij} - a_i a_j \E{2\s} \rb
 \end{bmatrix}, \qquad
G^{\mu\nu} =
 \begin{bmatrix}
	(-\E{-2\s} + a^2)	& -a^j \\
	-a^i	& g^{ij}
 \end{bmatrix},
}
where time redefinition invariance requires that $a^i$ is an independent gauge field, named as Kaluza-Klein gauge field. Using the euclidean time period $\tilde\b$ we can define the local equilibrium temperature of the theory as: $\vq_o = 1/\b_o = \E{-\s}/\tilde \b$. Our higher dimensional metric is hence disintegrated in a scalar (Temperature), a gauge field and a lower dimensional metric.

We can now use the metric $G^{\mu\nu}$ to raise/lower the components of vectors:
\bea{
	\cV_i = g_{ij}\cV^j + a_i\cV_0 , &\qquad \cV^0 = -\E{-2\s} \cV_0 - a_j\cV^j.
}
which are not Kaluza-Klein gauge invariant. From here we read out the $(d)$-covectors:
\bee{
	V_i = \lb \cV_i -a_i\cV_0 \rb.
}

Determinant of metric in two spaces can be related as:
\bee{
	G = - \det G_{\mu\nu} = \E{2\s} \det g_{ij} = \E{2\s} g.
}

We have the Levi-Civita symbol in lower spatial dimensions:
\bee{
	\e^{ijk\ldots} = \E{\s} \e^{0ijk\ldots} = -\E{-\s}\e_0^{\ ijk\ldots},
}
where $\e^{0123\ldots} = 1/\sqrt{G}$ and $\e^{123\ldots}= 1/\sqrt{g}$.

It is useful to see how higher dimensional contractions behave in lower dimensions:
\bee{
	\cA^\mu \cB_\mu
	= 
	-\E{-2\s} AB + A^i B_i
}
\bee{
	\e^{i\mu_1\mu_2\ldots\mu_{n-1}}\cA_{\mu_1\mu_2\ldots\mu_{n-1}},
	=
	\E{-\s} \e^{ij_1j_2\ldots j_{n-2}} \sum_a (-1)^a A_{j_1\ldots j_{a-1}0j_a\ldots j_{n-2}}.
}

\subsection{Derivatives of Metric}

Once the metric is known we can reduce the derivatives of metric, i.e. the Christoffel Symbol and the Riemann Tensor. The Christoffel Symbol is defined by:
\bee{
	\hat\G^{\l}_{\ \mu\nu} = \half G^{\l\a}\lb \dow_\mu G_{\a\nu} + \dow_\nu G_{\a\mu} - \dow_\a G_{\mu\nu} \rb.
}

Pretending it to be a tensor at the moment, if we define its indices to be raised and lowered with the metric $G^{\mu\nu}$. We can reduce it for Kaluza-Klein form of the metric:
\bee{\nn
	\hat\G_{000} = 0, \qquad
	\hat\G^i_{\ 00} = -\E{2\s}\frac{\dow^i\vq_o}{\vq_o}, \qquad
	\hat\G_{0 \ 0}^{\ i} = \hat\G_{00}^{\ \ i} = \E{2\s}\frac{\dow^i\vq_o}{\vq_o},
}
\bee{
	\hat\G^{ij}_{\ \ 0} = \hat\G^{i \ j}_{\ 0} = \half \E{2\s} f^{ij}, \qquad
	\hat\G_{0} ^{\ ij} = 
	- \half \E{2\s}g^{ia}g^{jb}\lb \dow_a a_b + \dow_b a_a \rb, \qquad
	\hat\G^{kij} =
	g^{il}g^{jm}\G^{k}_{\ lm},
}

where $\G^{k}_{\ ij}$ is Christoffel Symbol on $\cM_{(d)}$,  which is raised and lowered by $g_{ij}$. Also we define KK field tensor:
\bee{
	f^{ij} = \N^i a^j - \N^j a^i.
}

$\hat\G_0^{\ ij}$ is not KK gauge invariant, even though it has time index down and spatial index up, which is the manifestation of $\hat\G$ not being a tensor.

Lets define the higher dimensional covariant derivative as $\hat\N$ and lower dimensional as $\N$, whereas the usual derivative is given by $\dow$. We can check that:
\bea{
	\tilde \N^i \cV^j &= \N^i V^j + \half f^{ij} V, \nn \\
	\hat \N^i \cV_0 &= \N^i V + \b_o V \N^i \vq_o + \half\E{2\s}f^{ij} V_j, \nn \\
	\hat \N_0 \cV^i &= V \frac{\N^i \vq_o}{\vq_o} + \half \E{2\s} f^{ij} V_j, \nn \\
	\hat \N_0 \cV_0 &= \E{2\s} V^i \frac{\N_i \vq_o}{\vq_o}, \label{E:Grad00}
}

similarly,
\bea{
	\hat\N^i \cV^{jk} &= \N^i \cV^{jk} + \half f^{ij} \cV_{0}^{\ k} + \half f^{ik} \cV_{\ 0}^{j}, \nn \\
	\hat\N^i \cV^{j}_{\ 0} &= \N^i \cV^{j}_{\ 0} + \half f^{ij} \cV_{00} + =\cV^{j}_{\ 0} \frac{\N^i \vq_o}{\vq_o} + \half\E{2\s}f^{i}_{\ k} \cV^{jk}, \nn \\
	\hat\N^i \cV_{00} &= \N^i \cV_{00} + 2 \cV_{00} \frac{\N^i \vq_o}{\vq_o} + \half\E{2\s}f^{i}_{\ k} \lb \cV_{0}^{\ k} + \cV_{\ 0}^{k} \rb,  \nn \\
	\hat\N_0 \cV^{ij} &=  \cV^i_{\ 0} \frac{\N^j \vq_o}{\vq_o} + \cV^{\ j}_{0} \frac{\N^i \vq_o}{\vq_o} + \half \E{2\s} f^{j}_{\ k} \cV^{ik} + \half \E{2\s} f^{i}_{\ k} \cV^{kj}, \nn \\
	\hat\N_0 \cV^{i}_{\ 0} &= \cV_{00} \frac{\N^i \vq_o}{\vq_o} + \half \E{2\s} f^{i}_{\ k} \cV^{k}_{\ 0} + \E{2\s} \cV^{ij} \frac{\N_j \vq_o}{\vq_o}, \nn \\
	\hat\N_0 \cV_{00} &= \E{2\s} \lb \cV^{i}_{\ 0} + \cV^{\ i}_{0} \rb \frac{\N_i \vq_o}{\vq_o}.
}

Finally the Riemann Curvature Tensor is defined using an arbitrary vector $X^\mu$ as:
\bee{
	\cR_{\mu\nu\r\s} X^\s  \
	=
	\half\lb \hat\N_{\mu}\hat\N_{\nu} - \hat\N_{\nu} \hat\N_{\mu}\rb X_\r ,
}
using which we can define:
\bee{
	\cR_{\mu\nu} = \cR_{\mu\a\nu}^{\ \ \ \ \a}, \qquad \cR = \cR^\a_{\ \a}.
}

Now a straight away computation will yield:
\bea{
	\cR
	&=
	R
	- 4\frac{1}{\vq_o^2}\N_i\vq_o\N^i\vq_o
	+ 2\frac{1}{\vq_o}\N^i\N_i\vq_o
	+ \frac{1}{4}\E{2\s}f^{ij}f_{ij}, \nn \\
	u^\mu u^\nu\cR_{\mu\nu}
	=
	\E{-2\s}\cR_{00}
	&=
	2\frac{1}{\vq_o^2}\N_i \vq_o\N^i \vq_o
	- \frac{1}{\vq_o}\N_i \N^i \vq_o
	+ \frac{1}{4}\E{2\s}f^{ij}f_{ij}, \nn \\
	u_\mu\cR^{i\mu}
	=
	\E{-\s}\cR^i_{\ 0}
	&=
	\E{\s} \half \lb
		\N_k f^{ki}
		+ \frac{3}{\vq_o} f^{ik}\N_k\vq_o
	\rb, \nn \\
	\cR^{ij}
	&=
	R^{ij}
	- 2\frac{1}{\vq_o^2}\N^i \vq_o\N^j\vq_o
	+ \frac{1}{\vq_o}\N^i\N^j\vq_o
	+ \half \E{2\s}f^i_{\ a}f^{ja}, \nn \\
	u^\a u^\b \cR^{i \ j}_{\ \a\ \b} = \E{-2\s} \cR^{i \ j}_{\ 0\ 0}
	&=
	2\frac{1}{\vq_o^2}\N^i\vq_o\N^j\vq_o
	- \frac{1}{\vq_o}\N^i \N^j\vq_o
	+ \frac{1}{4}\E{2\s}f^i_{\ a}f^{ja}, \nn \\
	\cR^{ijk\a}u_\a = \E{-\s}\cR^{ijk}_{\ \ \ 0}
	&=
	\frac{1}{2\vq_o} \N^k f_1^{ij}
	- \frac{1}{\vq_o^2} \lb
		f_1^{ij}\N^k\vq_o 
		+ \half f_1^{ik} \N^j \vq_o
		- \half f_1^{jk} \N^i \vq_o
	\rb.
}

Here $R^{ijkl}$ is defined to be lower dimensional Riemann tensor, and $R^{ij}=R^{ikj}_{\ \ \ k}$, $R=R^{i}_{\ i}$.

\subsection{Derivatives of Gauge Field}

Now let us have a look at derivatives of gauge field $\cA^{\mu}$. Being a vector it decomposes as:
\bee{
	A = \cA_0 = -\E{2\s} \lb \cA^0 + a_jA^j \rb, \qquad
	A^i = \cA^i, \qquad
	A_i = \lb \cA_j -a_j A \rb.
}
The gauge transformation $\cA_\mu \ra \cA_\mu + \dow_\mu \L$ translates to:
\bee{
	A \ra A, \qquad
	A_i \ra A_i + \dow_i \L.
}

Hence $A^i$ is a gauge field on $\cM_{(d)}$, while $A$ is a scalar. Using $\tilde\b$ (euclidean temperature) we define the local equilibrium potential $\nu_o = \tilde \b A$. Higher dimensional field tensor however decomposes as:
\bee{
	\cF^{\mu\nu} = \hat\N^\mu \cA^\nu - \hat\N^\nu \cA^\mu
	\quad\Ra\quad 
	\cF^{ij} = F^{ij} + \E{\s}\vq_o\nu_o f^{ij}, \qquad
	\cF^{i}_{\ 0} = \E{\s}\vq_o\N^i \nu_o,
}
where,
\bee{
	F^{ij} = \N^i A^j - \N^j A^i.
}

Now we define the four vector electric field:
\bee{
	\cE^\mu = \cF^{\mu\nu} u_\nu 
	\quad\Ra\quad 
	\cE_0 = - \E{\s}\vq_o v_i \N^i \nu_o, \qquad
	\cE^i = - \E{-2\s} v \E{\s}\vq_o\N^i \nu_o + v_j \lb F^{ij} + \E{\s}\vq_o\nu_o f^{ij}\rb.
}

\section{Subsubleading Order Fluid} \label{apn:subsubleading}

In this appendix we extend the counting discussed in \cref{sec:counting} to subsubleading order fluid. We form a complete set of data at this order and classify the respective scalars, vectors and symmetric tensors. Later using the independent scalars at this order we construct an equilibrium partition function and compute its variation. We were however unable to process the constraints explicitly, as the calculations are analytically intractable.

\subsection{Counting at Equilibrium}

At subsubleading order ($D=n+1$, $s=3$), index families required are: $2D = 2n+2$  ($\bV_\e^{CC}$), $2D-1 =  2n+1$ ($\bT_\e^C$), $2D-2=2n$ ($\bV_\e^C$), $2D-3 =  2n-1$ ($\bT_\e$) and $2D-4=2n-2$ ($\bV_\e$). We only compute terms surviving at equilibrium, as non-equilibrium pieces are not required till subsubsubleading order parity-even or subsubsubsubleading order parity-odd calculation.

\paragraph{2D Family:}

$2D$ family was already discussed in \cref{sec:leadingcounting}, but this time since four indices are free from $\e$, two $(2,4,2)$ can appear with two antisymmetric indices of $R^{ijkl}$ contracted. We will find 3 combinations -- $(19n - 20)$ vectors of type $\bV^{CC}_\e$:
\begin{enumerate}
	\item $2(2,4,2) \oplus (n-3)(2,2,1)$: 1 possibility -- $(n-2)$ vectors
	\bc
		$\nfrac{m-1}{n-m-2}^{ijklm} \cK_{jk}^{\ \ ab} \cK_{lmab} \Big\vert_{m=1}^{n-2}$.
	\ec
	
	\item $(2,4,2) \oplus (n-1)(2,2,1)$ -- 4 possibilities -- $(8n-10)$ vectors
	\bc
		$\nfrac{m-1}{n-m}^{i}\cK \Big\vert_{m=1}^{n}$,
		$\nfrac{m-1}{n-m-1}^{ijk}\cX_{\L ja}\cK^{a}_{\ k} \Big\vert_{m=1}^{n-1}$,
		$\nfrac{m-1}{n-m-1}^{ijk}\cK_{jk}^{\ \ ab} \cX_{\L ab} \Big\vert_{m=1}^{n-1}$,
		$\nfrac{m-1}{n-m-2}^{ijklm}\cK_{jk}^{\ \ ab} \cX_{\L al} \cX_{\G bm} \Big\vert_{m=1}^{n-2}$.
	\ec
	
	\item $(n+1)(2,2,1)$ -- 3 possibilities -- $(10n-8)$ vectors
	\bc
		$\nfrac{m-1}{n-m}^{i} \cX^{jk}_\L \cX_{\G jk} \Big\vert_{m=1}^{n}$,
		$\nfrac{m-1}{n-m-1}^{ijk} \cX_{\L ja} \cX_{\G kb} \cX^{ab}_\Sigma \Big\vert_{m=1}^{n-1}$,
		$\nfrac{m-1}{n-m-2}^{ijklm} \cX_{1aj} \cX^a_{2 k} \cX_{1bl} \cX^b_{2 m} \Big\vert_{m=1}^{n-2}$.
	\ec
\end{enumerate}

\paragraph{2D-1 Family:}

$2D-1$ family was already discussed in \cref{sec:subleadingcounting}, but this time three indices are free from $\e$. So only one among $(2,4,2)$ and $(\frac{5}{3},5,3)$ can appear, and not more that once. We will find 5 combinations of type $\bT_\e^C$:
	\begin{enumerate}
		\item $(2,4,2) \oplus (n-2)(2,2,1) \oplus (1,1,1)$: 3 possibilities -- $(8n-12)$ symmetric traceless tensors
		\bc
			$\nfrac{m-1}{n-m-1}^{\langle ijk}\dow_j \vq_\L \cK_{k}^{\ a\rangle} \Big\vert_{m=1}^{n-1}$,
			$\nfrac{m-1}{n-m-1}^{\langle ijk} \cK^{\ \ a\rangle b}_{jk} \dow_{b} \vq_\L \Big\vert_{m=1}^{n-1}$,
			$\nfrac{m-1}{n-m-2}^{\langle ijklm} \dow_{j} \vq_\L \cX_{\G kb} \cK^{\ \ a\rangle b}_{lm} \Big\vert_{m=1}^{n-2}$.
		\ec
		
		\item $(n)(2,2,1) \oplus (1,1,1)$: 5 possibilities
		\begin{enumerate}
			\item Contraction between $(2,2,1)$ and $(1,1,1)$ -- $(12n-8)$ symmetric traceless tensors and $(4n)$ scalars
			\bc
				$\nfrac{m-1}{n-m}^{\langle i} \cX^{j\rangle k}_\L \dow_k\vq_{\G} \Big\vert_{m=1}^{n}$,
				$\nfrac{m-1}{n-m-1}^{\langle ijk} \cX^{l\rangle }_{\L j} \cX_{\G ka} \dow^a \vq_{\Sigma} \Big\vert_{m=1}^{n-1}$.
			\ec
			
			\textbf{Scalars:} We can take trace and get $4n$ scalars:
			\bc
				$\nfrac{m-1}{n-m}_{i} \cX^{ik}_\L \dow_k\vq_{\G} \Big\vert_{m=1}^{n}$
			\ec
			
			\item Contraction between $(2,2,1)$ and $(2,2,1)$ -- $(14n-18)$ traceless symmetric tensors and $(2n-2)$ scalars
			\bc
				$\nfrac{m-1}{n-m-1}^{\langle ijk} \cX^a_{1 j} \cX_{2 ak} \dow^{l\rangle }\vq_{\L} \Big\vert_{m=1}^{n-1}$.
				$\nfrac{m-1}{n-m-1}^{\langle ijk} \cX_{\L ja} \cX^{l\rangle a}_{\G} \dow_k \vq_{\Sigma} \Big\vert_{m=1}^{n-1}$,
				$\nfrac{m-1}{n-m-2}^{\langle ijklm} \cX^a_{1 j} \cX_{2 ak} \cX^{n\rangle }_{\L l} \dow_{m}\vq_{\G} \Big\vert_{m=1}^{n-2}$.
			\ec
			
			\textbf{Scalars:} Taking trace we get $2n-2$ scalars:
			\bc
				$\nfrac{m-1}{n-m-1}^{ijk} \cX^a_{1 j} \cX_{2 ak} \dow_i\vq_{\L} \Big\vert_{m=1}^{n-1}$.
			\ec
		\end{enumerate}
		
		\item $(2,4,2)\oplus(n-3)(2,2,1) \oplus (\frac{3}{2},3,2)$: 1 possibility -- $(2n-4)$ traceless symmetric tensors
		\bc
			$\nfrac{m-1}{n-m-2}^{\langle ijklm} \cK^{\ \ a\rangle b}_{jk} \dow_b\cX_{\L lm} \Big\vert_{m=1}^{n-2}$.
		\ec
		
		\item $(n-1)(2,2,1) \oplus (\frac{3}{2},3,2)$: 7 possibilities
		\begin{enumerate}
			\item Contraction within $(\frac{3}{2},3,2)$ -- $(6n-4)$ traceless symmetric tensors and $(2n)$ scalars
			\bc
				$\nfrac{m-1}{n-m}^{\langle i} \dow_k \cX^{kj\rangle }_\L \Big\vert_{m=1}^{n}$,
				$\nfrac{m-1}{n-m-1}^{\langle ijk} \cX^{a\rangle }_{\L j} \dow^b \cX_{\G bk} \Big\vert_{m=1}^{n-1}$.
			\ec
			
			\textbf{Scalars:} Taking trace we get $2n$ scalars:
			\bc
				$\nfrac{m-1}{n-m}^{i} \dow^k \cX_{\L ki} \Big\vert_{m=1}^{n}$.
			\ec
			
			\item Contraction between $(2,2,1)$ and $(\frac{3}{2},3,2)$ -- $(20n-28)$ traceless symmetric tensors and $(4n-4)$ scalars.
			\bc
				$\nfrac{m-1}{n-m-1}^{\langle ijk} \cX_{\L jb} \dow^b \cX_{\G k}^{a\rangle } \Big\vert_{m=1}^{n-1}$,
				$\nfrac{m-1}{n-m-1}^{\langle ijk} \cX_{\L jb} \dow^{a\rangle } \cX_{\G k}^{b} \Big\vert_{m=1}^{n-1}$,
				$\nfrac{m-1}{n-m-1}^{\langle ijk} \cX^{a\rangle b}_{\L} \dow_b \cX_{\G jk} \Big\vert_{m=1}^{n-1}$,
				$\nfrac{m-1}{n-m-2}^{\langle ijklm} \cX^{a\rangle }_{\L j} \cX_{\G kb} \dow^b \cX_{\Sigma lm} \Big\vert_{m=1}^{n-2}$.
			\ec
			
			\textbf{Scalars:} Taking trace we get $4n-4$ scalars:
			\bc
				$\nfrac{m-1}{n-m-1}^{ijk} \cX_{\L ib} \dow^b \cX_{\G jk} \Big\vert_{m=1}^{n-1}$.
			\ec
			
			\item Contraction between $(2,2,1)$ and $(2,2,1)$ -- $(2n-4)$ traceless symmetric tensors
			\bc
				$\nfrac{m-1}{n-m-2}^{\langle ijklm} \cX^{b}_{1 j} \cX_{2 bk} \dow^{a\rangle } \cX_{\L lm} \Big\vert_{m=1}^{n-2}$.
			\ec
			
		\end{enumerate}

		\item $(n-2)(2,2,1) \oplus (\frac{5}{3},5,3)$: 3 possibilities -- $(4n-6)$ traceless symmetric tensors
		\bc
			$\nfrac{m-1}{n-m-1}^{\langle ijk} \dow_a\cK_{\ \ jk}^{a l\rangle } \Big\vert_{m=1}^{n-1}$,
			$\nfrac{m-1}{n-m-1}^{\langle ijk} \dow_j\cK_{k}^{\ a\rangle } \Big\vert_{m=1}^{n-1}$,
			$\nfrac{m-1}{n-m-2}^{\langle ijklm} \dow_j\cK_{kl}^{\ \ a\rangle b}\cX_{\L bm} \Big\vert_{m=1}^{n-2}$.
		\ec
		
	\end{enumerate}

\paragraph{2D-2 Family:}

$2D-2$ family was already discussed in \cref{sec:subleadingcounting}. Here again, one among $(2,4,2)$ and $(\frac{5}{3},5,3)$ can appear, and not more that once. We will find 7 combinations -- $(39n - 46)$ vectors of type $\bV_\e^C$:
	\begin{enumerate}
		\item $(2,4,2)\oplus(D-4)(2,2,1)\oplus 2(1,1,1)$: No combinations possible
		\item $(n-1)(2,2,1)\oplus 2(1,1,1)$: 3 possibilities -- $(12n - 10)$ vectors
		\bc
			$\nfrac{m-1}{n-m}^{i} \dow_k \vq_{\L} \dow^k \vq_{\G} \Big\vert_{m=1}^{n}$,
			$\nfrac{m-1}{n-m-1}^{ijk}\cX_{\L ja} \dow^a \vq_{\G} \dow_k \vq_{\Sigma} \Big\vert_{m=1}^{n-1}$,
			$\nfrac{m-1}{n-m-2}^{ijklm}\cX^a_{1 j}\cX_{2 ak} \dow_l \vq_{1} \dow_m \vq_{2} \Big\vert_{m=1}^{n-2}$.
		\ec
		
		\item $(n-1)(2,2,1)\oplus (1,2,2)$: 2 possibilities -- $(6n - 4)$ vectors
		\bc
			$\nfrac{m-1}{n-m}^{i}\dow_k\dow^k \vq_\L \Big\vert_{m=1}^{n}$,
			$\nfrac{m-1}{n-1-m}^{ijk}\cX_{\L ja}\dow^a\dow_k \vq_\G \Big\vert_{m=1}^{n-1}$.
		\ec
		
		\item $(n-2)(2,2,1)\oplus (\frac{3}{2},3,2) \oplus (1,1,1)$: 3 possibilities -- $(16n - 24)$ vectors
		\bc
			$\nfrac{m-1}{n-m-1}^{ijk}\dow_j\vq_{\L} \dow^a \cX_{\G ak} \Big\vert_{m=1}^{n-1}$,
			$\nfrac{m-1}{n-m-1}^{ijk} \dow_{a} \cX_{\L jk}\dow^a \vq_{\G} \Big\vert_{m=1}^{n-1}$,
			$\nfrac{m-1}{n-m-2}^{ijklm} \cX_{\L ja} \dow^{a} \cX_{\G kl}\dow_m \vq_{\Sigma} \Big\vert_{m=1}^{n-2}$.
		\ec
		
		\item $(n-2)(2,2,1)\oplus (\frac{4}{3},4,3)$: 1 possibility -- $(2n-2)$ vectors
		\bc
			$\nfrac{m-1}{n-m-1}^{ijk}\dow_a\dow^a \cX_{\G jk} \Big\vert_{m=1}^{n-1}$.
		\ec
		
		\item $(n-3)(2,2,1)\oplus (\frac{5}{3},5,3) \oplus (1,1,1)$: No possiblities
		\item $(n-3)(2,2,1)\oplus 2(\frac{3}{2},3,2)$: 1 possibility -- $(3n-6)$ vectors
		\bc
			$\nfrac{m-1}{n-m-2}^{ijklm} \dow_{a} \cX_{\L jk} \dow^{a} \cX_{\G lm} \Big\vert_{m=1}^{n-2}$.
		\ec
	\end{enumerate}

\paragraph{2D-3 Family:} 

We are interested in combinations in $(2D-3)$ family which survive at equilibrium. We generated them through a Mathematica code and found 22 of them. We won't list all of them here, because it won't be required. Due to properties of $\bT_\e$, most of them will not contribute. We will be only left with 3 combinations -- $(7n-9)$ symmetric traceless tensors:
	\begin{enumerate}
		\item $(n-2)(2,2,1)\oplus 3(1,1,1)$: 1 possibility -- $(2n-2)$ symmetric traceless tensors
		\bc
			$\nfrac{m-1}{n-m-1}^{\langle ijk}\dow_j \vq_1 \dow_k \vq_2 \dow^{l\rangle } \vq_\L \Big\vert_{m=1}^{n-1}$.
		\ec
		\item $(n-2)(2,2,1)\oplus (1,1,1) \oplus (1,2,2)$: 1 possibility -- $(4n-4)$ symmetric traceless tensors
		\bc
			$\nfrac{m-1}{n-m-1}^{\langle ijk}\dow_j \vq_\L \dow_k \dow^{l\rangle } \vq_\G \Big\vert_{m=1}^{n-1}$.
		\ec
		\item $(n-3)(2,2,1)\oplus (\frac{3}{2},3,2) \oplus 2(1,1,1)$: 1 possibility -- $(2n-4)$ symmetric traceless tensors
		\bc
			$\nfrac{m-1}{n-m-2}^{\langle ijklm}\dow_j \vq_1 \dow_k \vq_2 \dow^{a\rangle } \cX_{\L lm} \Big\vert_{m=1}^{n-2}$.
		\ec
	\end{enumerate}
	
\paragraph{2D-4 Family:} There are 51 combinations in $(2D-4)$ family which survive at equilibrium. However none of them will contribute due to properties of $\bV_\e$.

All the subsubleading parity-odd data surviving at equilibrium has been summarized in \cref{tab:subsubleading_V,tab:subsubleading_T,tab:subsubleading_S}.

\mktbl{t}{tab:subsubleading_V}{Subsubleading Order Parity-odd Vectors at Equilibrium}{}
{|r|c|c|}
{
	\hline
	Name		& Term	& Equilibrium \\
	\hline
	\hline
	$\mathbb{\tilde V}_{1\L m}^{\mu} \big\vert_{m=1}^n$
	& $l^\mu \mathbf S_{1\L}$
	& $\nfrac{m-1}{n-m}^{i} \N^k\N_k \vq_{\L o}$ \\
	\hline
	$\mathbb{\tilde V}_{2(\L\G)m}^{\mu} \big\vert_{m=1}^n$
	& $l^\mu \mathbf S_{2(\L\G)}$
	& $\nfrac{m-1}{n-m}^{i} \N^k \vq_{\L o} \N_k \vq_{\G o}$ \\
	\hline
	$\mathbb{\tilde V}_{3(\L\G)m}^{\mu} \big\vert_{m=1}^n$
	& $l^\mu \mathbf S_{3(\L\G)}$
	& $\nfrac{m-1}{n-m}^{i} f^{ab}_\L f_{\G ab}$ \\
	\hline
	$\mathbb{\tilde V}_{4m}^{\mu} \big\vert_{m=1}^n$
	& $l^\mu \mathbf S_{4}$
	& $\nfrac{m-1}{n-m}^{i} R$ \\
	\hline
	$\mathbb{\tilde V}_{5\L\G m}^{\mu} \big\vert_{m=1}^{n-1}$
	& $\nfrac{m-1}{n-m-1}^{\mu\nu\r\s}u_\nu V_{\L\r} \mathbf V_{1\G\s}$
	& $\nfrac{m-1}{n-m-1}^{ijk} \N_j \vq_{\L o} \N^a f_{\G ak}$ \\
	\hline
	$\mathbb{\tilde V}_{6\L\G\Sigma m}^{\mu} \big\vert_{m=1}^{n-1}$
	& $\nfrac{m-1}{n-m-1}^{\mu\nu\r\s}u_\nu V_{\L\r} \mathbf V_{2\G\Sigma\s}$
	& $\nfrac{m-1}{n-m-1}^{ijk} \N_j \vq_{\L o} f_{\G ka}\N^a \vq_{\Sigma o} $ \\
	\hline
	$\mathbb{\tilde V}_{7\L\G m}^{\mu} \big\vert_{m=1}^{n-1}$
	& $\nfrac{m-1}{n-1-m}^{\mu\nu\r\s}u_\nu\cX_{\L \r\a} \hat\N^\a\hat\N_\s \vq_\G$
	& $\nfrac{m-1}{n-1-m}^{ijk} f_{\L ja} \N^a\N_k \vq_{\G o}$ \\
	\hline
	$\mathbb{\tilde V}_{8\L m}^{\mu} \big\vert_{m=1}^{n-1}$
	& $\nfrac{m-1}{n-1-m}^{\mu\nu\r\s}u_\nu\cX_{\L \r\a} \cK^\a_{\ \s}$
	& $\nfrac{m-1}{n-1-m}^{ijk} f_{\L ja} R^a_{\ k}$ \\
	\hline
	$\mathbb{\tilde V}_{9(\L\G)\Sigma m}^{\mu} \big\vert_{m=1}^{n-1}$
	& $\nfrac{m-1}{n-m-1}^{\mu\nu\r\s} u_\nu \cX_{\L\r\a} \cX_{\G\s\b} \cX^{\a\b}_\Sigma$
	& $\nfrac{m-1}{n-m-1}^{ijk} f_{\L ja} f_{\G kb} f^{ab}_\Sigma$ \\
	\hline
	$\mathbb{\tilde V}_{10\L\G m}^{\mu} \big\vert_{m=1}^{n-1}$
	& $\nfrac{m-1}{n-m-1}^{\mu\nu\r\s}u_\nu V^\a_{\L} \hat\N_{\a}\cX_{\G\r\s}$
	& $\nfrac{m-1}{n-m-1}^{ijk} \N^a \vq_{\L o} \N_{a}f_{\G jk}$ \\
	\hline
	$\mathbb{\tilde V}_{11\L m}^{\mu} \big\vert_{m=1}^{n-1}$
	& $\nfrac{m-1}{n-m-1}^{\mu\nu\r\s} u_\nu P^{\a\b}\hat\N_\a\hat\N_\b \cX_{\L \r\s}$
	& $\nfrac{m-1}{n-m-1}^{ijk} \N^a\N_a f_{\L jk}$ \\
	\hline
	$\mathbb{\tilde V}_{12\L m}^{\mu} \big\vert_{m=1}^{n-1}$
	& $\nfrac{m-1}{n-m-1}^{\mu\nu\r\s} u_\nu \cK_{\r\s}^{\ \ \a\b} \cX_{\L \a\b}$
	& $\nfrac{m-1}{n-m-1}^{ijk} R_{jk}^{\ \ ab} f_{\L ab}$ \\
	\hline
	$\mathbb{\tilde V}_{13 m}^{\mu} \big\vert_{m=1}^{n-2}$
	& $\nfrac{m-1}{n-m-2}^{\mu\nu\r\s\a\b}u_\nu \cX_{1\k\r} \cX^\k_{2 \s} \hat\N_\a \vq_{1} \hat\N_\b \vq_{2}$
	& $\nfrac{m-1}{n-m-2}^{ijklm} f_{1 aj} f^a_{2 k} \N_l \vq_{1o} \N_m \vq_{2o}$ \\
	\hline
	$\mathbb{\tilde V}_{14 m}^{\mu} \big\vert_{m=1}^{n-2}$
	& $\nfrac{m-1}{n-m-2}^{\mu\nu\r\s\a\b} u_\nu \cX_{1\k\r} \cX^\k_{2 \s} \cX_{1\d\a} \cX^\d_{2 \b}$
	& $\nfrac{m-1}{n-m-2}^{ijklm} f_{1aj} f^a_{2k} f_{1bl} f^b_{2m}$ \\
	\hline
	$\mathbb{\tilde V}_{15 \L\G\Sigma m}^{\mu} \big\vert_{m=1}^{n-2}$
	& $\nfrac{m-1}{n-m-2}^{\mu\nu\r\s\a\b} u_\nu \cX_{\L \k\r} \hat\N^{\k} \cX_{\G \s\a} V_{\Sigma\b}$
	& $\nfrac{m-1}{n-m-2}^{ijklm} f_{\L aj} \N^{a} f_{\G kl} \N_m \vq_{\Sigma o}$ \\
	\hline
	$\mathbb{\tilde V}_{16 (\L\G) m}^{\mu} \big\vert_{m=1}^{n-2}$
	& $\nfrac{m-1}{n-m-2}^{\mu\nu\r\s\a\b} u_\nu P^{\d\k}\hat\N_{\d} \cX_{\L \r\s} \hat\N_{\k} \cX_{\G \a\b}$
	& $\nfrac{m-1}{n-m-2}^{ijklm} \N_{a} f_{\L jk} \N^{a} f_{\G lm}$ \\
	\hline
	$\mathbb{\tilde V}_{17(\L\G)m}^{\mu} \big\vert_{m=1}^{n-2}$
	& $\nfrac{m-1}{n-m-2}^{\mu\nu\r\s\a\b} u_\nu \cK_{\r\s}^{\ \ \k\d} \cX_{\L \k\a} \cX_{\G \d\b}$
	& $\nfrac{m-1}{n-m-2}^{ijklm} R_{jk}^{\ \ ab} f_{\L al} f_{\G bm}$ \\
	\hline
	$\mathbb{\tilde V}_{18 m}^{\mu} \big\vert_{m=1}^{n-2}$
	& $\nfrac{m-1}{n-m-2}^{\mu\nu\r\s\a\b} u_\nu \cK_{\r\s}^{\ \ \d\k} \cK_{\a\b\d\k}$
	& $\nfrac{m-1}{n-m-2}^{ijklm} R_{jk}^{\ \ ab} R_{lmab}$ \\
	\hline
}

\mktbl{t}{tab:subsubleading_T}{\centering Subsubleading Order Parity-odd Symmetric Traceless Tensors at Equilibrium}{}
{|r|c|c|}
{
	\hline
	Name	& Term	& Equilibrium \\
	\hline
	\hline
	$\mathbb{\tilde T}_{1\L m}^{\mu\nu} \big\vert_{m=1}^{n}$
	& $l_m^{\langle \mu} \mathbf V^{\nu\rangle}_{1\L}$
	& $\nfrac{m-1}{n-m}^{\langle i} \N_k f^{kj\rangle}_\L$ \\
	\hline
	$\mathbb{\tilde T}_{2\L m}^{\mu\nu} \big\vert_{m=1}^{n}$
	& $l_m^{\langle \mu} \mathbf V^{\nu\rangle}_{2\L\G}$
	& $\nfrac{m-1}{n-m}^{\langle i} f^{j\rangle k}_\L \N_k\vq_{\G}$ \\
	\hline
	$\mathbb{\tilde T}_{3\L m}^{\mu\nu} \big\vert_{m=1}^{n-1}$
	& $\mathbf {\tilde V}_{1m}^{\langle \mu} V^{\nu\rangle}_\L$
	& $\nfrac{m-1}{n-m-1}^{\langle ijk} f^a_{1 j} f_{2 ak} \N^{l\rangle}\vq_{\L}$ \\
	\hline
	$\mathbb{\tilde T}_{4\L m}^{\mu\nu} \big\vert_{m=1}^{n-1}$
	& $\mathbf {\tilde V}_{2m}^{\langle\mu} V^{\nu\rangle}_\L$
	& $\nfrac{m-1}{n-m-1}^{\langle ijk} \N_j\vq_{1o}\N_k\vq_{2o} \N^{a\rangle} \vq_{\L o}$ \\
	\hline
	$\mathbb{\tilde T}_{5\L\G m}^{\mu\nu} \big\vert_{m=1}^{n-1}$
	& $\nfrac{m-1}{n-m-1}^{\langle \mu\nu\r\s} u_\nu V_{\L\r} \hat\N_\s \hat\N^{\a \rangle} \vq_\G$
	& $\nfrac{m-1}{n-m-1}^{\langle ijk} \N_j \vq_{\L o} \N_k \N^{a\rangle} \vq_{\G o}$ \\
	\hline
	$\mathbb{\tilde T}_{6\L m}^{\mu\nu} \big\vert_{m=1}^{n-1}$
	& $\nfrac{m-1}{n-m-1}^{\langle\mu\nu\r\s}u_\nu V_{\L\r} \cK_{\s}^{\ \a\rangle}$ 
	& $\nfrac{m-1}{n-m-1}^{\langle ijk} \N_j \vq_{\L o} R_{k}^{\ a\rangle}$  \\
	\hline
	$\mathbb{\tilde T}_{7\L\G m}^{\mu\nu} \big\vert_{m=1}^{n-1}$
	& $\nfrac{m-1}{n-m-1}^{\langle\mu\nu\r\s} u_\nu \cX^{\a\rangle}_{\L \r} \mathbf V_{1\G\s}$
	& $\nfrac{m-1}{n-m-1}^{\langle ijk} f^{a\rangle}_{\L j} \N^b f_{\G bk}$ \\
	\hline
	$\mathbb{\tilde T}_{8\L\G\Sigma m}^{\mu\nu} \big\vert_{m=1}^{n-1}$
	& $\nfrac{m-1}{n-m-1}^{\langle\mu\nu\r\s} u_\nu \cX^{\a\rangle}_{\L \r} \mathbf V_{2\G\Sigma\s}$
	& $\nfrac{m-1}{n-m-1}^{\langle ijk} f^{a\rangle}_{\L j} f_{\G kb}\N^b \vq_{\Sigma o}$ \\
	\hline
	$\mathbb{\tilde T}_{9\L\G\Sigma m}^{\mu\nu} \big\vert_{m=1}^{n-1}$
	& $\nfrac{m-1}{n-m-1}^{\langle\mu\nu\r\s} u_\nu \cX_{\L \r\a} \cX^{\d\rangle\a}_{\G} V_{\Sigma\s}$
	& $\nfrac{m-1}{n-m-1}^{\langle ijk} f_{\L jb} f^{a\rangle b}_{\G} \N_k \vq_{\Sigma o}$ \\
	\hline
	$\mathbb{\tilde T}_{10\L\G m}^{\mu\nu} \big\vert_{m=1}^{n-1}$
	& $\nfrac{m-1}{n-m-1}^{\langle\mu\nu\r\s} u_\nu \cX_{\L \r\d} \hat\N^{\a\rangle} \cX_{\G \s}^{\d}$
	& $\nfrac{m-1}{n-m-1}^{\langle ijk} f_{\L jb} \N^{a\rangle} f_{\G k}^{b}$ \\
	\hline
	$\mathbb{\tilde T}_{11\L\G m}^{\mu\nu} \big\vert_{m=1}^{n-1}$
	& $\nfrac{m-1}{n-m-1}^{\langle\mu\nu\r\s} u_\nu \cX_{\L \r\d} \hat\N^\d \cX_{\G \s}^{\a\rangle}$
	& $\nfrac{m-1}{n-m-1}^{\langle ijk} f_{\L jb} \N^b f_{\G k}^{a\rangle}$ \\
	\hline
	$\mathbb{\tilde T}_{12\L\G m}^{\mu\nu} \big\vert_{m=1}^{n-1}$
	& $\nfrac{m-1}{n-m-1}^{\langle\mu\nu\r\s} u_\nu \cX^{\a\rangle\d}_{\L} \hat\N_\d \cX_{\G \r\s}$
	& $\nfrac{m-1}{n-m-1}^{\langle ijk} f^{a\rangle b}_{\L} \N_b f_{\G jk}$ \\
	\hline
	$\mathbb{\tilde T}_{13\L m}^{\mu\nu} \big\vert_{m=1}^{n-1}$
	& $\nfrac{m-1}{n-m-1}^{\langle\mu\nu\r\s}u_\nu V_{\L\a} \cK^{\ \ \k\rangle\a}_{\r\s}$
	& $\nfrac{m-1}{n-m-1}^{\langle ijk} \N_b\vq_{\L o} R^{\ \ a\rangle b}_{jk}$ \\
	\hline
	$\mathbb{\tilde T}_{14m}^{\mu\nu} \big\vert_{m=1}^{n-1}$
	& $\nfrac{m-1}{n-m-1}^{\langle\mu\nu\r\s} u_\nu \hat\N_\a\cK_{\r\s}^{\ \ \k\rangle\a}$
	& $\nfrac{m-1}{n-m-1}^{\langle ijk} \N_b R_{jk}^{\ \ a\rangle b}$ \\
	\hline
	$\mathbb{\tilde T}_{15m}^{\mu\nu} \big\vert_{m=1}^{n-1}$
	& $\nfrac{m-1}{n-m-1}^{\langle\mu\nu\r\s} u_\nu \hat\N_\r\cK_{\s}^{\ \a\rangle}$
	& $\nfrac{m-1}{n-m-1}^{\langle ijk} \N_j R_{k}^{\ a\rangle}$ \\
	\hline
	$\mathbb{\tilde T}_{16\L m}^{\mu\nu} \big\vert_{m=1}^{n-2}$
	& $\nfrac{m-1}{n-m-2}^{\langle\mu\nu\r\s\a\b}u_\nu\hat\N_\r \vq_1 \hat\N_\s \vq_2 \hat\N^{\k\rangle} \cX_{\L \a\b}$
	& $\nfrac{m-1}{n-m-2}^{\langle ijklm } \N_j \vq_{1o} \N_k \vq_{2o} \N^{a\rangle} f_{\L lm}$ \\
	\hline
	$\mathbb{\tilde T}_{17\L m}^{\mu\nu} \big\vert_{m=1}^{n-2}$
	& $\nfrac{m-1}{n-m-2}^{\langle\mu\nu\r\s\a\b} u_\nu \cX^{\d}_{1 \r} \cX_{2 \d\s} \hat\N^{\k\rangle} \cX_{\L \a\b}$
	& $\nfrac{m-1}{n-m-2}^{\langle ijklm} f^{b}_{1 j} f_{2bk} \N^{a\rangle} f_{\L lm}$ \\
	\hline
	$\mathbb{\tilde T}_{18\L\G m}^{\mu\nu} \big\vert_{m=1}^{n-2}$
	& $\nfrac{m-1}{n-m-2}^{\langle\mu\nu\r\s\a\b} u_\nu \cX^\d_{1 \r} \cX_{2 \d\s} \cX^{\k\rangle}_{\L \a} V_{\G\b}$
	& $\nfrac{m-1}{n-m-2}^{\langle ijklm} f^b_{1 j} f_{2bk} f^{a\rangle}_{\L l} \N_m\vq_{\G o}$ \\
	\hline
	$\mathbb{\tilde T}_{19\L\G\Sigma m}^{\mu\nu} \big\vert_{m=1}^{n-2}$
	& $\nfrac{m-1}{n-m-2}^{\langle\mu\nu\r\s\a\b} u_\nu \cX^{\k\rangle}_{\L \r} \cX^\d_{\G \s} \hat\N_\d \cX_{\Sigma \a\b}$
	& $\nfrac{m-1}{n-m-2}^{\langle ijklm} f^{a\rangle}_{\L j} f^b_{\G k} \N_b f_{\Sigma lm}$ \\
	\hline
	$\mathbb{\tilde T}_{20\L\G m}^{\mu\nu} \big\vert_{m=1}^{n-2}$
	& $\nfrac{m-1}{n-m-2}^{\langle\mu\nu\r\s\a\b} u_\nu V_{\L\r} \cX_{\G \s \d} \cK^{\ \ \k\rangle\d}_{\a\b}$
	& $\nfrac{m-1}{n-m-2}^{\langle ijklm} \N_j \vq_{\L o} f_{\G kb} R^{\ \ a\rangle b}_{lm}$ \\
	\hline
	$\mathbb{\tilde T}_{21\L m}^{\mu\nu} \big\vert_{m=1}^{n-2}$
	& $\nfrac{m-1}{n-m-2}^{\langle\mu\nu\r\s\a\b} u_\nu \hat\N_\d\cX_{\L \r\s} \cK^{\ \ \k\rangle\d}_{\a\b}$
	& $\nfrac{m-1}{n-m-2}^{\langle ijklm} \N_b f_{\L jk} R^{\ \ a\rangle b}_{lm}$ \\
	\hline
	$\mathbb{\tilde T}_{22\L m}^{\mu\nu} \big\vert_{m=1}^{n-2}$
	& $\nfrac{m-1}{n-m-2}^{\langle\mu\nu\r\s\a\b} u_\nu \cX_{\L \d\r} \hat\N_\s\cK_{\a\b}^{\ \ \k\rangle\d}$
	& $\nfrac{m-1}{n-m-2}^{\langle ijklm} f_{\L bj} \N_k R_{lm}^{\ \ a\rangle b}$ \\
	\hline
}

\mktbl{t}{tab:subsubleading_S}{\centering Subsubleading Order Parity-odd Scalars at Equilibrium}{}
{|r|c|c|}
{
	\hline
	$\mathbb{\tilde S}_{1\L m} \big\vert_{m=1}^{n-1}$
	& $\mathbf {\tilde V}_{1m}^{\mu} V_{\L \mu}$
	& $\nfrac{m-1}{n-m-1}^{ijk} f_{1ai} f^{a}_{2 j} \N_k \vq_{\L o}$ \\
	\hline
	$\mathbb{\tilde S}_{2\L\G m} \big\vert_{m=1}^{n-1}$
	& $\nfrac{m-1}{n-m-1}^{\mu\nu\r\s} u_\nu \cX_{\L\mu\d} \hat\N^\d \cX_{\G \r\s}$
	& $\nfrac{m-1}{n-m-1}^{ijk} f_{\L ia} \N^a f_{\G jk}$ \\
	\hline
	$\mathbb{\tilde S}_{3\L m} \big\vert_{m=1}^n$
	& $l^\mu_{m} \mathbf V_{1\L\mu}$
	& $\nfrac{m-1}{n-m}^{i} \N^k f_{\L ki}$ \\
	\hline
	$\mathbb{\tilde S}_{4\L\G m} \big\vert_{m=1}^n$
	& $l^\mu_{m} \mathbf V_{2\L\G\mu}$
	& $\nfrac{m-1}{n-m}^{i} f_{\L ij}\N^j\vq_{\G o}$ \\
	\hline
}

\subsubsection*{Independent Scalars}

As we discussed in \cref{sec:basisdata}, we only need to construct independent scalars which enter in equilibrium partition function. At subsubleading order one can find antisymmetrizations which will determine $\bbS_{3\L m}$ and $\bbS_{4\L\G m}$ in terms of $\bbS_{1\L m}$ and $\bbS_{2\L\G m}$ respectively:
\bee{
	\cX_1^{[i_1j_1}\ldots \cX_1^{i_{m-1}j_{m-1}}\cX_2^{i_mj_m}\ldots \cX_2^{i_{n-1}j_{n-1}} \cX^a_{\L b} \hat\N^{b]}\vq_\G \Big\vert_{m=1}^n = 0,
}
\bee{
	\cX_1^{[i_1j_1}\ldots \cX_1^{i_{m-1}j_{m-1}}\cX_2^{i_mj_m}\ldots \cX_2^{i_{n-1}j_{n-1}} \hat\N_{b} \cX^{ba]}_{\L} \Big\vert_{m=1}^n = 0.
}

Each of $\bbS_{1\L m}$ and $\bbS_{2\L\G m}$, on the other hand is a unique scalar per choice of the parity-even tensor used to construct it by contracting with $\e$. Since antisymmetrization conditions cannot alter the tensor structure, these scalars are independent.

\subsection{Attempt for Fluid Constraints}

In the equilibrium partition function $\tilde\D^{(n+1)}W_{(C)}^{eqb}$ we can include the scalars: $\tilde{\mathbb S}_{o1\L m}$, $\tilde{\mathbb S}_{o2\L\G m}$. But it can be checked that antisymmetric part of $\tilde{\mathbb S}_{o2[\L\G] m}$ can be related through a total derivative to $\tilde{\mathbb S}_{o1\L m}$. So we have:
\bee{
	\tilde\D^{(n+1)} W_{(C)}^{eqb} =
	\int \lbr x^i \rbr\sqrt{g} \ ^{n-2}\bC_{m-1} \lbr
		\cQ_{1,\L m}\tilde{\mathbb S}_{o1,\L m}
		+ \cQ_{2,\L\G m}\tilde{\mathbb S}_{o2,(\L\G) m}
	\rbr.
}

Sum over relevant indices is understood. Varying the partition function we will get:
\bem{
	\frac{\d \tilde\D^{(n+1)} W_{(C)}^{eqb}}{\d g_{ij}} 
	= 
	\ ^{n-2}\bC_{m-1} \lB
		\cQ_{1,\L m} \lb
			\tilde\bbT^{ij}_{o9,12\L m}
			+ \tilde\bbT^{ij}_{o8,21\L m}
		\rb
		+ 2 \cQ_{2,\L\G m} \tilde\bbT^{ij}_{o7,\L\G m}
	\rB \\
	- \ ^{n-3}\bC_{m-1} (n-2) \lB
		\cQ_{1,\L m} \tilde\bbT^{ij}_{o18,2\L m}
		- \cQ_{2,\L\G (m+2-\Sigma)} \tilde\bbT^{ij}_{o19,\L\Sigma\G m}
	\rB,
}
\bee{
	\frac{\d \tilde\D^{(n+1)}W_{(C)}^{eqb}}{\d \vq_{o\L}} =
	\ ^{n-2}\bC_{m-1} \lb 
		2 \dow_{[\L} \cQ_{1,\G] m} \tilde{\mathbb S}_{o1,\G m}
		- \cQ_{1,\L m} \tilde{\mathbb S}_{o2,[12] m}
		+ \dow_\L\cQ_{2,\G\Sigma m}\tilde{\mathbb S}_{o2,(\G\Sigma) m}
	\rb,
}
\bem{
	\frac{\d \tilde\D^{(n+1)}W_{(C)}^{eqb}}{\d a_{\O i}} =
	\ ^{n-2}\bC_{m-1} \lB
		(-)^\O \dow_\G\cQ_{1,\L m} \tilde{\mathbb V}^i_{o6,\G \bar\O \L m}
		- (-)^\O \cQ_{1,\L m} \lb
			\tilde{\mathbb V}^i_{o7,\bar\O\L m}
			+ \half \tilde{\mathbb V}^i_{o10,\L\bar\O m}
		\rb \dbrk
		- 2 \dow_\Sigma\dow_\G\cQ_{2,\O\L m} \tilde{\mathbb V}^i_{o6,\Sigma\L\G m}
		+ \dow_\G\cQ_{2,\O\L m} \lb
			4 \tilde{\mathbb V}^i_{o5,\G\L m}
			+ 2 \tilde{\mathbb V}^i_{o7,\L\G m}
			+ \tilde{\mathbb V}^i_{o10,\G\L m}
		\rb \dbrk
		+ 2 \cQ_{2,\O\L m} \lb
			2 \tilde{\mathbb V}^i_{o8,\L m}
			+ \tilde{\mathbb V}^i_{o11,\L m}
			+ \tilde{\mathbb V}^i_{o12,\L m}
		\rb
	\rB \\
	+ (n-2) \ ^{n-3}\bC_{m-1} \lB
		- 2\lb
			\dow_\G \cQ_{2,\O\L (m+2-\Sigma)}
			- \half\dow_\G\cQ_{2,\L\Sigma (m+2-\O)}
		\rb \tilde{\mathbb V}^i_{o15,(\Sigma\L)\G m} \dbrk
		+ \lb
			\cQ_{2,\O\L (m+2-\Sigma)} 
			- \half  \cQ_{2,\L\Sigma (m+2-\O)}
		\rb \lb
			\tilde{\mathbb V}^i_{o16,(\Sigma\L) m}
			- 2 \tilde{\mathbb V}^i_{o17,(\Sigma\L) m}
		\rb
	\rB
}

On the other hand from counting we can see that the third order parity-odd corrections (at eqb.) are given by (see \cref{tab:subsubleading_V,tab:subsubleading_T,tab:subsubleading_S}):
\bee{\label{E:-3cons}
	\tilde\U^\mu_{o(n+1)} = \sum_{\#} \f_{o\#} \mathbb{\tilde V}_{o\#}^{\mu}, \qquad
	\tilde\Pi^{\mu\nu}_{o(n+1)} = 
	\sum_{\#} \vp_{o\#} \mathbb{\tilde T}_{o\#}^{\mu\nu}
	+ P^{\mu\nu}_o \sum_{\#} \g_{o\#} \mathbb{\tilde S}_{o\#}.
}
$\#$ corresponds to all the relevant indices. Similar to subleading order, here also we will have special contributions for $n=2$, as 3 leading order ($n-1$) parity odd corrections can combine to give a $3n-3$ order parity-odd corrections, which will be equal to $n+1$ only at $n=2$ (Remember we are not considering $n=1$ case). Now comparing \cref{E:CurrentsGeneric} with \cref{E:parti_perturb} at parity-odd subsubleading derivative order, we have corrections to constitutive relations:
\bea{
	\frac{1}{\vq_o}\tilde\pi^{ij}_{o(n+1)} 
	&= 
	2 \frac{\d \tilde\D^{(n+1)}W_{(C)}^{eqb}}{\d g_{ij}} 
	- g^{ij}\bA_{o\L} \frac{ \d \tilde\D^{(n+1)} W_{(C)}^{eqb}}{\d \vq_{o\L}}
	+ g^{ij} \frac{2}{\vq_o} \bA_{o\L} \dow_\L P_o \tilde\D^{(n-1)}v^{k} \D^{(2)} v_k \nn \\
	& \qquad + g^{ij} \bA_{o2} \tilde\D^{(n-1)} v_i \lb
		\vs^i_{o(2)}
		+ \boxed{ \tilde\D^{(1)} \bar\vs^i_{(1)}}
	\rb
	- g^{ij} \bA_{o2} \D^{(2)} v_k \lb
		q_o \tilde\D^{(n-1)}v^{k}
		- \tilde\vs^k_{o(n-1)}
	\rb \nn \\
	& \qquad - 2 P_o^{(1,0)}\tilde\D^{(n-1)}v^{(i} \D^{(2)} v^{j)}
	- \frac{1}{\vq_o}\lb
		\tilde\D^{(n-1)}\pi^{ij}_{(2)} 
		+ \tilde\D^{(n)}\pi^{ij}_{(1)}
	\rb, \nn \\
	P^{(1,0)}_o\tilde\vs^{i}_{o(n+1)}
	&= 
	\vq_o \bS_{o\L}\frac{ \d \tilde\D^{(n+1)} W_{(C)}^{eqb}}{\d a_{\L i}}
	- \lB
		P_o^{(1,0)} \D^{(2)}q 
		- \frac{1}{\vq_o^2} P^{(0,1)}_o  \D^{(2)}(\e+P) 
	\rB \tilde\D^{(n-1)}v^i \nn \\
	&\qquad + q_o \tilde\D^{(n-1)}v_j \lb
		\boxed{\frac{3}{2} P_o^{(1,0)} \tilde\D^{(1)} v^i \tilde\D^{(1)}v^j}
		+ \frac{1}{\vq_o}\pi^{ij}_{o(2)}
	\rb \nn \\
	& \qquad - P^{(1,0)}_o \lb
		\D^{(2)}\tilde\vs^{i}_{(n-1)}
		+ \tilde\D^{(n-1)}\vs^{i}_{(2)}
		+ \tilde\D^{(n)}\vs^{i}_{(1)}
	\rb, \label{E:subsuboddidentifications}
}

while the fluid variables get corrections:
\bea{
	\tilde\D^{(n+1)}v^i
	&= 
	(-)^\L \frac{\mu_{o\L}}{P^{(1,0)}_o} \frac{ \d \tilde\D^{(n+1)} W_{(C)}^{eqb}}{\d a_{\L i}}
	- \frac{1}{P^{(1,0)}_o}\tilde\D^{(n-1)}v_j \lb
		\frac{1}{\vq_o}\D^{(2)} (\e+P) g^{ij}
		+ \frac{1}{\vq_o}\pi^{ij}_{o(2)}
	\rb \nn \\
	& \qquad - \boxed{\frac{3}{2} \tilde\D^{(1)} v^i \tilde\D^{(1)}v^j \tilde\D^{(1)}v_j}, \nn \\
	\tilde\D^{(n+1)}\vq_{\L}
	&= 
	\vq_o \bE_{o\L\G} \frac{ \d \tilde\D^{(n+1)} W_{(C)}^{eqb}}{\d \vq_{o\G}}
	- 2 \bA_{o\L} \tilde\D^{(n-1)}v^{k} \D^{(2)} v_k \nn \\
	&\qquad - \bE_{o\L 2} \vq_o \tilde\D^{(n-1)} v_i \lb
		\vs^i_{o(2)}
		+ \boxed{ \tilde\D^{(1)} \bar\vs^i_{(1)}}
	\rb
	+ \bE_{o\L 2} \vq_o \D^{(2)} v_i \lb
		q_o \tilde\D^{(n-1)} v^i
		- \tilde\vs^i_{o(n-1)}
	\rb. \label{E:subsuboddcorrections}
}

From here onwards in principle the way would be to solve
\cref{E:subsuboddidentifications} and find constraints for transport
coefficients appearing in \cref{E:-3cons}. To solve we would need to
plug in the fluid variable corrections to all lower orders, along with
corrections to lower order constitutive relations due to fluid
variable corrections. The terms which were zero at equilibrium at
lower orders will also start to contribute by gaining the fluid
variable corrections. Leaving aside terms specifically for $n=2$,
still we would have to deal with a large mess in
\cref{E:subsuboddidentifications} which is analytically not quite
tractable. So we leave these expressions at this point for reference.

Readers are advised that expressions
\cref{E:subsuboddidentifications,E:subsuboddcorrections} does not
contain contributions from gravitational and mixed anomaly, and
conserved Chern Simons form. Recall that while we set up relations
\cref{E:parti_perturb}, we only used the form of anomalous currents
\cref{E:anomalouscurrent} and conserved Chern-Simons form
\cref{E:lowerCS} to subleading derivative order. At subsubleading
order, they will receive further gravitational corrections.

\end{document}